\documentclass[usenatbib]{mn2e}

\usepackage{graphicx}
\usepackage[usenames]{color}

\newlength{\colwidth}
\newcommand{\Msolyrkpcsq}{\mbox{M}_{\sun}\,\mbox{yr}^{-1}\,\mbox{kpc}^{-2}}
\newcommand{\Msolpcsq}{\mbox{M}_{\sun}\,\mbox{pc}^{-2}}
\newcommand{\Msolh}{h^{-1}\,\mbox{M}_{\odot}}
\newcommand{\Myr}{\mbox{Myr}}
\newcommand{\kpch}{h^{-1}\,\mbox{kpc}}
\newcommand{\pch}{h^{-1}\,\mbox{pc}}

\newcommand{\K}{\mbox{K}}

\newcommand{\ion}[2]{\hbox{{\sc #1}\,{\sc #2}}}
\newcommand{\HI}{\ion{H}{I}} 

\newcommand{\cm}{{\rm cm}}

\newcommand{\kms}{{\rm km}\,{\rm s}^{-1}}

\newcommand{\yr}{{\rm yr}}

\newcommand{\Msun}{{{\rm M}_\odot}}

\begin{document}

\title[Schmidt and Kennicutt-Schmidt laws]{On the relation between
  the Schmidt and Kennicutt-Schmidt star formation laws and its
  implications for numerical simulations}

\author[J. Schaye \& C. Dalla Vecchia]
{Joop Schaye\thanks{E-mail: schaye@strw.leidenuniv.nl} and
Claudio Dalla Vecchia\thanks{E-mail: caius@strw.leidenuniv.nl}\\
Leiden Observatory, Leiden University, P.O. Box 9513, 2300 RA Leiden,
  the Netherlands
}

\maketitle

\abstract 
When averaged over large scales, star formation in galaxies is
observed to follow the empirical Kennicutt-Schmidt (KS) law for surface
densities above a constant threshold. While the observed
law involves surface densities, theoretical models and simulations
generally work with volume density laws (i.e.\ Schmidt laws). We
derive analytic relations between star formation laws expressed in
terms of surface densities, volume densities, and pressures and we
show how these relations depend on parameters such as the effective
equation of state of the multiphase interstellar medium. Our analytic
relations enable us to implement observed surface density laws into
simulations. Because the parameters of our prescription for star formation
are observables, we are not free to tune them to match the observations. 
We test our theoretical framework using high-resolution simulations of
isolated disc galaxies that assume an effective equation of state for the
multiphase interstellar medium. We are able to reproduce the star
formation threshold and both the slope and the normalisation of
arbitrary input KS laws without tuning any parameters and with very
little scatter, even for unstable galaxies and even if we use poor
numerical resolution. Moreover, we can do so for arbitrary effective
equations of state. Our prescription therefore enables simulations of
galaxies to bypass our current inability to simulate the formation of
stars. On the other hand, the fact that we can reproduce arbitrary
input thresholds and KS laws, rather than just the particular 
ones picked out by nature, indicates that simulations that
lack the physics and/or resolution to simulate the multiphase
interstellar medium can only provide limited insight into the origin of
the observed star formation laws.
\endabstract


\section{Introduction}
\label{sec:intro}
The efficiency of star formation (SF) and its environmental dependence are
fundamental, but poorly understood, ingredients for models of the
formation and evolution of galaxies. Fortunately, for many purposes,
the models do not require a 
detailed understanding of SF in order
to make progress. Our lack of understanding can often be bypassed by
using empirical scaling
relations which give the efficiency of SF averaged over
scales that are large compared to those of individual star
clusters. Such scaling relations can be built into semi-analytic
models and numerical simulations of galaxy formation, allowing the
SF process to be treated as a black box calibrated to
reproduce the observations.

Observationally, a combination of two SF laws is known to
describe the efficiency of SF reasonably well for nearby
galaxies \citep[e.g.][]{Kennicutt1998review}: a SF threshold and
a Kennicutt-Schmidt (KS) law:
\begin{equation}
\dot{\Sigma}_\ast = \left \{
\begin{array}{lr} 
0 & \mbox{ if } \Sigma_{\rm g} < \Sigma_{\rm c} \\
A \left (\Sigma_{\rm g}/1~\Msolpcsq \right )^n & \mbox{ if } \Sigma_{\rm g} \ge \Sigma_{\rm c}
\end{array}
\right .
\label{eq:genSFlaw}
\end{equation}
where $\dot{\Sigma}_\ast$ is the rate of SF per
unit area and per unit time, $\Sigma_{\rm g}$ is the gas surface density,
and $\Sigma_{\rm c}$ is the threshold surface density for SF. It is
important to distinguish the empirical KS law, $\dot{\Sigma}_\ast
\propto \Sigma_{\rm g}^n$ from the \cite{Schmidt1959} law, $\dot{\rho}_\ast
\propto \rho_{\rm g}^{n_S}$, which is often assumed in theoretical
models. We will reserve $n$ exclusively for KS (i.e.\ surface density)
laws and will use $n_S$ to denote the power-law index of Schmidt
(i.e.\ volume density) laws.

The existence of a surface density threshold for SF has
long been explained in terms of the \cite{Toomre1964} criterion for
instability
\citep[e.g.][]{Quirk1972,Kennicutt1989,Martin&Kennicutt2001},
 usually under the assumption of a constant velocity 
dispersion. This explanation was, however, criticized by 
\citet[][hereafter S04; see also Schaye 2007]{Schaye2004}, who argued
that the 2-D Toomre analysis is not 
applicable to the warm interstellar medium (ISM) in the outer parts of
disc galaxies and that it 
fares less well in comparison to observations than a simple
constant surface density threshold, even allowing for an ad-hoc
rescaling of the critical Toomre Q-parameter, as in
\cite{Kennicutt1989} and subsequent studies.  

The fact that a constant 
threshold of $\Sigma_{\rm c} \sim 10~\Msolpcsq$ works well empirically, had
been known for some time 
\citep[e.g.][]{Guiderdoni1987,Skillman1987}. \cite{Elmegreen&Parravano1994}
emphasized that the need for a cold gas phase imposes a pressure
threshold. S04 showed that the transition from the warm to
the cold gas phase triggers gravitational instability on a wide range
of scales. He demonstrated that this thermo-gravitational instability
can account for the observed SF threshold, which he
predicted to fall in the range $\Sigma_{\rm c} \approx 3$--$10~\Msolpcsq$
(corresponding to a critical pressure $P_{\rm c}/k \sim
10^2$--$10^3~\cm^{-3}\,\K$ and a critical volume density $n_{\rm H}
\sim 10^{-2}$--$10^{-1}~\cm^{-3}$), with weak dependencies on the
metallicity, UV radiation field, turbulent pressure, and the mass
fraction in gas. Conversely, below the threshold the presence of
ionizing radiation keeps
the gas warm ($T\sim 10^4~\K$) and stable. The model can therefore
also account for the 
observed constant velocity dispersion of $\approx 8~\kms$ in the outer
parts of disc galaxies \citep[e.g.][]{Meurer1996}. Several recent
studies have found that the 
S04 criterion can account for high-resolution observations of SF in
nearby disc galaxies 
\citep[e.g.][]{Deblok&Walter2006,Auld2006} and of the formation
of star clusters in tidal arms \citep{Maybate2007}. The study by
\cite{Deblok&Walter2006} is particularly relevant because it avoids
azimuthal smoothing, which is inappropriate for local thresholds and
which has often confused comparisons between theories of local SF
thresholds and observation. 

For $\Sigma_{\rm g} \gg \Sigma_{\rm c}$, many studies have found that the
simple KS law, with $n\approx 1$--2, can well represent
the global, disc-averaged star formation rate (SFR) over at least 5 decades
in $\Sigma_{\rm g}$ (see \citealt{Kennicutt1998review} for a
review). \cite{Kennicutt1998} found\footnote{The normalization assumes
  a Salpeter initial mass function over 0.1--$100~\Msun$. Note that
  the error bars on 
  the normalisation and slope of the \cite{Kennicutt1998} result must
  be strongly correlated, since all his data points are on one side of
  the pivot point of $1~\Msolpcsq$. Changing one parameter by itself,
  will therefore lead to results that are inconsistent with the
  observations.}
\begin{equation}
\dot{\Sigma}_\ast = (2.5 \pm 0.7) \times 10^{-4} \,\Msolyrkpcsq \left
    ({\Sigma_{\rm g} 
    \over 1~\Msolpcsq}\right )^{(1.4 \pm 0.15)}
\label{eq:KS}
\end{equation}
as the best fit for his sample of 97 galaxies spanning 5 orders of
magnitude in $\Sigma_{\rm g}$, ranging from quiescently star-forming disc
galaxies to starbursts. 
The KS law can also be expressed in terms of the
gas consumption time-scale
\begin{eqnarray}
t_{\rm g} &\equiv& {\Sigma_{\rm g} \over \dot{\Sigma}_\ast}
= A^{-1} \left (1~\Msolpcsq \right )^n \Sigma_{\rm g}^{1-n} \label{eq:tgas} \\
&=& (4 \pm 1)\times 10^9 ~\yr
\left ({\Sigma_{\rm g} \over 1~\Msolpcsq}\right )^{-(0.4 \pm 0.15)}.
\label{eq:tgas_nrs}
\end{eqnarray}
The same law is also found to be a good
description of the relation between the SFR and
gas surface density when they are azimuthally averaged in radial bins
\citep[e.g.][]{Zhang2001,Wong&Blitz2002,Komugi2005,Boissier2006,Schuster2007}
and even when they are measured in local cylindrical bins, although in
that case the coefficient $A$ may be somewhat smaller
\citep{Kennicutt2007}. 

Note that the agreement between the azimuthally
averaged and globally averaged (i.e., averaged
over an entire galaxy) SF laws is surprising. Since the
observed KS law is non-linear (i.e.\ $n \ne 1$), the global law should
only equal the local one if individual galaxies have constant surface
densities. The fact that the same law works in both cases therefore
implies that, 
in terms of their SF properties, galaxies have effectively
constant surface densities. This will be approximately true if the
global SFR is dominated by the nucleus. 

The KS law depends on the total gas surface density, but several
studies have claimed that the exponent $n\approx 1.4$ is only valid
for molecular gas and that including atomic gas results in larger
exponents \citep[e.g.][]{Wong&Blitz2002,Heyer2004}. However, the \HI\
surface density is observed to 
saturate at $\Sigma_{\rm HI} \sim 10~\Msolpcsq$, which can be
explained as the result of the formation of a cold gas phase in clouds
with surface densities that exceed this critical value
\citep{Schaye2001maxNHI}. Such clouds will become unstable and will
eventually become mostly molecular, unless the process is halted by
SF and associated feedback processes. Since, not
coincidentally (S04), this surface density is similar to 
the threshold for SF, \HI\ will contribute
significantly to (or even dominate) the gas surface density for
densities below or around 
the threshold. It is therefore not surprising that studies focusing on
this low density regime, where a significant fraction of the gas at a
fixed radius will have a surface density below the SF threshold, find
steeper KS laws.  Empirical
prescriptions that take into account the molecular fraction may work
better in this regime \citep{Blitz&Rosolowsky2006}. This does, however, not
undermine the validity of the KS law for $\Sigma_{\rm g} \gg \Sigma_{\rm c}$.

Many theories have been proposed to explain the observed KS law
\cite[e.g.][]{Silk1997,Tan2000,Elmegreen2002,Krumholz&McKee2005,Tutukov2006}. 
Most rely in some way  
or another on the assumption that the SFR is
proportional to the gas density over the dynamical time-scale,
\begin{equation}
\dot{\rho}_\ast = \epsilon {\rho_{\rm g} \over t_{\rm dyn}} \sim \epsilon
G^{1/2} \rho_{\rm g}^{3/2},
\label{eq:Schmidt_dyn}
\end{equation}
where $\epsilon$ is the SF efficiency per dynamical
time. If $\epsilon$ is independent of the density, then the result
is a Schmidt law with exponent $n_S=1.5$. If, moreover, the
scale height ($H 
\sim \Sigma_{\rm g}/\rho_{\rm g}$) is also constant, then we have a KS law with
exponent 1.5, which agrees with the empirically determined value at
the 1$\sigma$ level. Note, however, that the assumption of a constant
scale height is non-trivial. We will show that this assumption is
unnecessary, generally incorrect, and that dropping it gives us
considerable insight as well as a much more accurate recipe for the
implementation of SF laws in numerical simulations. 

Numerical simulations of the formation of galaxies
\cite[e.g.][]{Cen&Ostriker1992,Summers1993,Navarro&White1993,Steinmetz&Mueller1994,Mihos&Hernquist1994,Katz1996,Gnedin1996,Gerritsen&Icke1997,Yepes1997,Springel2000,Thacker&Couchman2000,Kay2002,Kawata&Gibson2003,Kravtsov2003,Springel&Hernquist2003,Sommer-Larsen2003,Marri&White2003,Okamoto2003,Tasker&Bryan2006},
as well as semi-analytic models \cite[e.g.][]{Kauffmann1993} generally
use a Schmidt law of the form (\ref{eq:Schmidt_dyn}), although other
types of models have been proposed \cite*[e.g.][]{Li2005,Booth2007} and
some models use different exponents for the Schmidt law
\cite[e.g.\ $n_S=1$;][]{Kravtsov2003}. The Schmidt law is usually combined
with a threshold volume density, $\rho_{\rm c}$, and an upper limit on the
allowed temperature for SF. Other requirements, such as
converging flows and Jeans instability are sometimes invoked, but
these are often sensitive to resolution and tend to be relatively
unimportant if the problem is numerically resolved. The
efficiency per dynamical time and the threshold volume density are the
main free parameters, which are tuned to make the simulations fit some
observable, e.g.\ the observed SF threshold and KS law for
the case of simulations of isolated galaxies. 

In the next section we will derive the
relation between the KS and Schmidt laws analytically. We will
show that the success of the simulations depends on
parameters that are generally ignored, such as the effective equation
of state of the multiphase gas and the gas fraction. In
section~\ref{sec:simrecipe} we will provide a
recipe to reproduce arbitrary KS laws for arbitrary equations of state and
gas fractions. Contrary to other prescriptions, the parameters of our
module cannot 
be tuned because their values are derived analytically from the
observed surface density threshold and KS law. In
section~\ref{sec:sims} we will then use 
high-resolution hydrodynamical simulations to show that our recipe
works extremely well. Despite the absence of tunable parameters, our 
prescription is able to reproduce observed 
SF laws with much higher precision than has been possible
in the past.

\section{Relating surface density, volume density, and pressure laws}
\label{sec:analytics}

The KS law involves surface densities, whereas the Schmidt law deals
with volume densities. Prescriptions for SF in
three-dimensional simulations of galaxy formation cannot be expressed
in terms of surface densities, but do aim to match the observed KS
law. Although 
related, there are important differences between volume and surface
density laws. In the literature it is often assumed that the exponents
of the corresponding KS and Schmidt power laws are the same, but, as
we shall see, this is generally not true. 

The KS law describes the SFR when averaged over scales that
are large compared with individual star clusters. In our study of its
relation with the Schmidt law, we will therefore assume that 
the volume densities, as well as the pressure, are averaged over
similar scales. This is precisely what is relevant for simulations of galaxy
formation which currently lack both the resolution and the physical
ingredients that are necessary to model the
multiphase ISM. 

\subsection{Thresholds}
\label{sec:analytics_thresholds}

Volume and surface densities can be related as follows 
\citep[][S04]{Schaye2001}. If self-gravity is important, the
density will typically fluctuate on the local Jeans scale. For the case of
self-gravitating discs, this implies that the scale height 
will be of the order the local Jeans scale. Hence, the gas column
density is of the order of the ``Jeans column density''
\begin{eqnarray}
\Sigma_{\rm g} & \sim & \Sigma_{\rm g,J} \equiv \rho_{\rm g} L_{\rm J}\\
&=& \left ({\gamma k \over \mu G X}\right )^{1/2} (f n_{\rm H}
  T)^{1/2} \label{eq:SigmaJ_nH} \\
&=& \left ({\gamma\over G}\right )^{1/2} (f_{\rm g} P_{\rm tot})^{1/2},
\label{eq:SigmaJ_P}
\end{eqnarray}
where $\gamma$ is the ratio of specific heats, $X$ the hydrogen mass
fraction, $f\equiv f_{\rm g}/f_{\rm th}$, with $f_{\rm g}$ the mass fraction in gas (within
a scale height of the gas) and
$f_{\rm th}$ the fraction of the midplane pressure that is thermal, and
$P_{\rm tot}$ is the total midplane pressure (i.e., including both thermal
and non-thermal components). Putting
in numbers yields
\begin{eqnarray}
\Sigma_{\rm g} & \approx & 29~\Msolpcsq ~f^{1/2} \left ({T \over 10^4~\K}\right
    )^{1/2} \left ({n_{\rm H} \over 
    1~\cm^{-3}}\right )^{1/2} \label{eq:SigmaJ_nH_nrs} \\
&\approx & 28~\Msolpcsq ~f_{\rm g}^{1/2} \left ({P_{\rm tot}/k \over
    10^4~\cm^{-3}\,\K}\right )^{1/2},
\label{eq:SigmaJ_P_nrs}
\end{eqnarray}
where we assumed $\gamma=5/3$, $\mu = 1.23$ and $X=0.752$.

If the gas
is far from local hydrostatic equilibrium, then the scale height
differs substantially from the Jeans length. Far out of
equilibrium means in this case that either $t_{\rm dyn} \ll t_{\rm
  sc}$ or  $t_{\rm dyn} \gg t_{\rm sc}$, where $t_{\rm sc} \equiv H
/c_s$ is the local sound crossing time, with $H$ the scale height and
$c_{s,{\rm eff}} = (\gamma P_{\rm tot}/\rho_{\rm g})^{1/2}$ the effective sound
speed.  If left undisturbed, the
gas will return to local equilibrium on the time-scale $\min(t_{\rm
  dyn},t_{\rm sc})$, which is very short for the densities of
interest here ($t_{\rm dyn} \sim 1 / \sqrt{G\rho} \approx 8\times 10^7 ~\yr
~f_{\rm g}^{1/2} \left (n_{\rm H} / 1~\cm^{-3}\right )^{-1/2}$).
Note that if the entire disc were far from hydrostatic
equilibrium, then the scale height would fluctuate strongly
everywhere, contrary to what is observed.

Following S04, equations (\ref{eq:SigmaJ_nH}) and
(\ref{eq:SigmaJ_P}) can be used 
to convert the observed surface density threshold for SF into
a threshold volume density or a threshold pressure. Assuming that
$f\sim f_{\rm g}\sim 1$ and $T\sim 10^4~\K$ at the threshold for
SF, and using his prediction for the surface density threshold,
$\Sigma_{\rm c} \sim 3$--$10~\Msolpcsq$, we obtain $n_{\rm H,c} \sim
10^{-2}$--$10^{-1}~\cm^{-3}$ or $P_{\rm tot,c}/k \sim
10^2$--$10^3~\cm^{-3}\,\K$. This agrees well with the recipes used in
numerical simulations, which typically set $n_{\rm H,c} \sim
10^{-1}~\cm^{-3}$, although some studies of individual galaxies have
used much higher values (e.g., $n_{\rm H,c} = 50~\cm^{-3}$ for
\citealt{Kravtsov2003} and $10^3~\cm^{-3}$ for \citealt{Li2005}). 

\subsection{KS laws}
\label{sec:analytics_KSlaws}

Using (\ref{eq:SigmaJ_P}) we can write the
gas consumption time-scale (\ref{eq:tgas}) as
\begin{equation} 
t_{\rm g} = A^{-1} \left (1~\Msolpcsq\right )^n \left ({\gamma \over G} f_{\rm g} P_{\rm tot}\right
  )^{(1-n)/2}.
\label{eq:tgas_P} 
\end{equation}
Hence, we can write the Schmidt law as
\begin{equation} 
\dot{\rho}_\ast \equiv  {\rho_{\rm g} \over t_{\rm g}}
= A \left (1~\Msolpcsq\right )^{-n} \left ({\gamma \over G} f_{\rm g} P_{\rm tot}\right
)^{(n-1)/2} \rho_{\rm g}.
\label{eq:Schmidt_inbetween}
\end{equation}
We could write this expression somewhat differently by making use of
the ideal gas law, eliminating
either $\rho_{\rm g}$ or $P_{\rm tot}$, but at the cost of introducing an
explicit dependence on $T$.
However, the Schmidt law can be further simplified if we assume that the
effective equation of state
of the multiphase ISM, when averaged over large scales, is polytropic
\begin{equation}
P_{\rm tot} = P_{\rm tot,c} \left ({\rho_{\rm g} \over \rho_{\rm g,c}}\right
    )^{\gamma_{\rm eff}},
\label{eq:eos}
\end{equation}
where $\gamma_{\rm eff}$ is the polytropic index, not to be
confused with the ratio of specific heats $\gamma$.
This assumption allows us to eliminate one variable and we can write,
\begin{eqnarray}
\dot{\rho}_\ast & =& A \left (1~\Msolpcsq\right )^{-n} \nonumber \\
&& \times \left ({\gamma \over G} f_{\rm g} P_{\rm c}
\right )^{(n-1)/2} \rho_{\rm g,c} \left ({\rho_{\rm g} \over \rho_{\rm g,c}}\right
)^{{(n-1)\gamma_{\rm eff} \over 2} + 1}.
\label{eq:Schmidt}
\end{eqnarray}

We can see from equation (\ref{eq:Schmidt}) that the power-law index
of the Schmidt law is in general not the same as the power-law index
of the corresponding KS law. For a polytropic equation of state we have
\begin{equation}
n_S = {(n-1)\gamma_{\rm eff} \over 2} + 1
\end{equation}
or, equivalently,
\begin{equation}
n = {2(n_S-1) \over \gamma_{\rm eff}} + 1.
\label{eq:n(n_S)}
\end{equation}
Hence, the two power-law indices are equal if $\gamma_{\rm
  eff}=2$ \citep[see also][]{Springel2000}, but for any other equation
of state equality requires $n=1=n_S$.

Simulations of galaxy formation that do not explicitly impose an
equation of state for gas with a density that exceeds the SF
threshold, typically predict that the star forming gas is isothermal,
$T(\rho_{\rm g}>\rho_{\rm c}) 
\sim 10^4~\K$, because the cooling function cuts off sharply at
$10^4~\K$ if molecular cooling is not included. Using $\gamma_{\rm
  eff}=1$ gives $n = 2$ for the commonly used Schmidt law index
$n_S=1.5$. To obtain the observed value of $n=1.4$, one would need to
use $n_S = 1.2$. To obtain $n=1.4$ using $n_S=1.5$, the simulations
would need to have $\gamma_{\rm eff} = 2.5$. 

Equation (\ref{eq:Schmidt}) shows that the Schmidt law depends also on
the gas fraction, albeit weakly: $\dot{\rho}_\ast \propto
f_{\rm g}^{(n-1)/2}$. For $n=1.4$ this gives $\dot{\rho}_\ast \propto
f_{\rm g}^{0.2}$. If the gas fraction depends on the gas density, then this
will change the relation between $n$ and $n_S$ somewhat. If the gas
fraction were to decrease with increasing density, then the effective value
of $n_S$ would be somewhat smaller (greater) for fixed $n>1$
($n<1$). Conversely, a given Schmidt law would correspond to a
somewhat steeper (shallower) KS law for $n_S > 1$ ($n_S < 1$).

\section{Application to simulations}
\label{sec:simrecipe}
The observed SF threshold, $\Sigma_{\rm c}$, can be converted into a threshold
for $n_{\rm H}$ or $P_{\rm tot}$ using equations
(\ref{eq:SigmaJ_nH}) and (\ref{eq:SigmaJ_P}), respectively. If the
local gas fraction is not known a priori, as will generally be the
case for ab initio simulations, then it is necessary to assume a
constant value. Similarly, the fraction of the pressure that is
thermal will generally also need to be fixed. Following
S04, we advise assuming $f_{\rm g} = f_{\rm th} = 1$. At the
threshold, in the outer parts of galaxies, the disc is mostly gaseous,
there is no SF to inject turbulent energy, and the gas is
kept warm by UV (background) radiation. In fact,
the thermal velocity dispersion can by itself account for the observed
line widths in the outer disc, leaving little room for turbulence
(S04). When using a volume density threshold, we therefore advise using
$T=10^4~\K$, appropriate for self-gravitating disc with $\Sigma_{\rm g} \sim
\Sigma_{\rm c}$ (S04). We expect the gas to be mostly in atomic form at the threshold
and therefore advise using 
the corresponding value of the mean particle weight $\mu$. 

Only gas with a temperature below some critical value $T_{\rm c}$ should be
allowed to form stars. For simulations that lack either the physics
(e.g.\ molecular cooling and radiative transfer) or the resolution to
model the formation of a cold gas phase, critical temperatures in the
range $10^4~\K \ll T_{\rm c} \la 10^5~\K$
are appropriate. An upper limit on the temperature is necessary to  
prevent hot gas from forming stars before it cools. This
is particularly important for models that employ a pressure threshold,
rather than a volume density threshold, because hot gas can exceed
such a threshold even for $\rho_{\rm g} \ll \rho_{\rm g,c}(T=10^4~\K)$. 
The results are insensitive to the exact value of $T_{\rm c}$, as
long as it is much greater than the value below which the cooling
time rises sharply ($\sim 10^4~\K$ for most simulations of galaxy
formation) and as long as it is not much greater than the value for
which the cooling time is minimum ($\sim 10^5~\K$). We use
$T_{\rm c}=10^5~\K$. 

Cosmological
simulations should also employ a threshold density contrast to avoid
spurious SF at very high redshift ($z\ga 10^2$), when the mean density
of the universe is similar to the proper density threshold. It is
common to choose a value similar to the density contrast at the virial
radius of a collapsed object, $\Delta_{\rm c} \equiv \rho_{\rm
  g}/\bar{\rho}_{\rm g}
\sim 60$. The  
precise value is unimportant, as long as $10 \ll \Delta_{\rm c} <
\rho_{\rm g,c}/\bar{\rho}_{\rm g}(z_{\rm c})$, where $z_{\rm c}$ is the redshift at which
the first resolved objects collapse in the simulation. 

The observed KS law, with its normalisation $A$ and slope $n$, can be
implemented as the corresponding Schmidt law, equation
(\ref{eq:Schmidt_inbetween}), or, if a polytropic equation of state is
assumed, equation (\ref{eq:Schmidt}). However, since
simulations require $\dot{m}_\ast$ rather than
$\dot{\rho}_\ast$, the SFR can be expressed as a function of the total
pressure, without any direct dependence on either density or temperature:
\begin{eqnarray}
\dot{m}_\ast & \equiv & {m_{\rm g} \over t_{\rm g}} = m_{\rm g} {\dot{\rho}_\ast \over
  \rho_{\rm g}} \\
&=& A \left (1~\Msolpcsq\right )^{-n} m_{\rm g} \left ({\gamma \over G} f_{\rm g} P_{\rm tot}\right
)^{(n-1)/2},
\label{eq:mstardot}
\end{eqnarray}
where $m_{\rm g}$ is the gas mass of the element\footnote{A gas element can
  either be a particle or a cell, depending on the type of
  simulation.} for which we want to
compute $\dot{m}_\ast$, and we used (\ref{eq:tgas_P}). This equation
can be converted to a function of $\rho_{\rm g}$, $T$ and $\mu$ using the
ideal gas law, or it can be rewritten as a function of $\rho_{\rm g}$ if a
monotonic equation of state is assumed. However, by implementing the
SF law in the form of equation (\ref{eq:mstardot}), the same
expression can be used regardless of whether an effective equation of
state is assumed.  

Putting in
numbers, we obtain 
\begin{equation} 
t_{\rm g} \approx 1.67\times 10^9~\yr \left ({f_{\rm g} P_{\rm tot}/k \over
  10^3~\cm^{-3}\,\K}\right )^{-0.2}
\end{equation}
and
\begin{equation}
\dot{m}_\ast 
= 5.99\times 10^{-10}~\Msun\,\yr^{-1} \left ({m_{\rm g} \over 1~\Msun}\right )
\left ({f_{\rm g} P_{\rm tot}/k \over 10^3~\cm^{-3}\,\K}\right
)^{0.2}
\label{eq:dotmstar}
\end{equation}
for the observed KS law (equation [\ref{eq:KS}]) and $\gamma=5/3$.
As for the
threshold, it will generally be necessary to assume a constant value
of $f$ (note that 
$f_{\rm g}P_{\rm tot} = f P_{\rm th}$). Fortunately, the dependence on $f$ is
very weak for the observed value of $n$ ($\dot{m}_\ast \propto
f^{0.2}$ for $n=1.4$).

Simulations must interpret the SF law stochastically. That is, the
probability that a gas element with SFR
$\dot{\rho}_\ast$ is converted into a star particle in a time step
$\Delta t$ is\footnote{Alternatively, one can follow \citet{Katz1992}
and write $Prob = 1 - e^{-\Delta t/t_{\rm g}}$, which cannot exceed unity
and reduces to (\ref{eq:prob}) for $\Delta t \ll t_{\rm g}$. What form is
preferable depends on the interpretation of individual resolution
elements. In our simulations the time step is always very small
compared to $t_{\rm g}$. Clearly, any simulation for which $\Delta t \ga
t_{\rm g}$ will produce spurious results.}
\begin{equation}
Prob = \min\left ({\Delta t \over t_{\rm g}},1\right ) = \min\left
({\dot{m}_\ast \Delta t \over  m_{\rm g}},1\right ), 
\label{eq:prob}
\end{equation}
where the gas consumption time-scale and the SFR are
given by equations (\ref{eq:tgas_P}) and (\ref{eq:mstardot}),
respectively. 
Since the masses of star particles typically exceed the mass of
individual stars 
by many orders of magnitude, they are to be interpreted as simple stellar
populations. Simulations that allow gas elements to spawn multiple
star particles, rather than
converting entire gas elements into star particles, should scale
the probability for SF accordingly: $P \rightarrow P' = P m_{\rm g}/m_\ast$. 

As an aside, we note that spawning multiple generations of star
particles per gas element, an approach frequently used in the
literature, can have
undesirable consequences. Suppose, for 
example, that star particles inject energy from core collapse
supernovae into the surrounding gas particles or, in the case of mesh
codes, into the nearest cell(s). This is a strategy that is
commonly used to mimic the effect of
feedback from SF. As $m_\ast/m_{\rm g}$ is decreased, the change
in the energy per
unit mass of the receiving gas elements, and therefore their
temperature increase, will be correspondingly reduced. Because the
cooling time is sensitive to the temperature, this will change the
efficiency of the feedback. Adding the energy in kinetic form does not
alleviate this problem: decreasing $m_\ast/m_{\rm g}$ will lead to lower
velocities and thus smaller post-shock temperatures. Kicking the
neighboring gas elements with a fixed velocity, but with a probability
proportional to the 
stellar mass formed also does not solve the problem. As $m_\ast/m_{\rm g}$
is decreased, the creation of star 
particles would be less and less likely to be accompanied by a kick,
allowing most of the SF events to proceed without any
feedback to accompany them.

\subsection{The effective equation of state}
\label{sec:simrecipe_eos}

Simulations that do not attempt to simulate star (cluster) formation
from first principles, i.e., simulations that lack the resolution
and/or the physics to model a multiphase ISM, will benefit from the
SF recipe presented here. However, since 
star forming gas (i.e.\ gas with $P_{\rm tot} > P_{\rm c}$ and $T < T_{\rm c}$) is
predicted to be 
multiphase, one cannot interpret the temperature predicted
by the simulation for this gas in the usual way. It therefore also
does not make sense to compute cooling rates, since
those depend on the actual kinetic temperature of the gas. Instead, it
is preferable to specify an effective equation of state a priori. Assuming a
specific equation of state makes explicit what can and what cannot be
simulated, does not waste resources on
calculations that are wrong, and also allows one to
prevent some numerical problems, as we shall see below. 

Although it is possible to devise semi-analytic sub-grid models that
predict the effective pressure of the multiphase ISM as a function of the
average density of the gas represented by a resolution element
\cite[e.g.][]{Yepes1997,Springel&Hernquist2003}, we prefer, for simplicity,
to use a polytropic equation of state, equation (\ref{eq:eos}). 

The choice of effective polytropic index $\gamma_{\rm eff}$, not to be
confused with the ratio of specific heats $\gamma$, can be
important. Larger values result in a steeper increase of the pressure
with the gas (volume) density and thus, for $n>1$, a shorter gas consumption
time-scale at a fixed density. This means that the maximum densities
in the simulation will become smaller and the stellar distribution
less concentrated. 

For a polytropic equation of state $P_{\rm tot} \propto
\rho_{\rm g}^{\gamma_{\rm eff}}$, the Jeans length and mass scale as
\begin{eqnarray}
L_{\rm J} & \propto & f_{\rm g}^{1/2} \rho_{\rm g}^{(\gamma_{\rm eff}-2)/2} \label{eq:L_J}
\\
M_{\rm J} & \propto & f_{\rm g}^{3/2} \rho_{\rm g}^{(3\gamma_{\rm eff}-4)/2}.
\label{eq:M_J}
\end{eqnarray}
Hence, $\gamma_{\rm eff}=4/3$ gives a Jeans mass independent of the
gas density and a Jeans length that scales as $L_{\rm J}\propto
\rho_{\rm g}^{-1/3}$. If we take
$\gamma_{\rm eff} = 2$, then the Jeans length will be constant, whereas
$M_{\rm J} \propto \rho_{\rm g}$. Hence, for $\gamma_{\rm eff} \ge 2$ we do not
expect self-gravity to promote further collapse, for $4/3 < \gamma_{\rm
  eff} < 2$ we expect gas clouds to collapse without fragmenting, and
for $\gamma_{\rm eff} < 4/3$ we expect both collapse and, possibly,
fragmentation.  

A Jeans mass that does not decrease with increasing
density is desirable from a numerical point of view. The
alternative, a Jeans mass that decreases with density, will 
lead to a situation in which particles have masses greater than the
local Jeans mass, which is known
to lead to artificial fragmentation \citep{Bate&Burkert1997}. A
polytropic index $\gamma_{\rm eff} = 4/3$ allows one to prevent the
strong artifacts that result from the inability to resolve the Jeans
mass, while at the same time allowing collapse to proceed. 

For Smoothed Particle Hydrodynamics (SPH) simulations 
$\gamma_{\rm eff} = 4/3$ not only fixes the ratio of the Jeans mass
and the mass 
resolution, it does the same for the ratio of the Jeans length and the spatial
resolution. Provided the gravitationally softening is chosen
sufficiently small, the spatial resolution scales with the SPH kernel
$h$ and thus as $h\propto (m_{\rm g}/\rho_{\rm g})^{1/3} \propto (m_{\rm g}/M_{\rm J})^{1/3}
L_{\rm J}$, where $m_{\rm g}$ is the particle mass.  Hence, if $m_{\rm g}/M_{\rm J}$ is
independent of the density, then so is $h/L_{\rm J}$. For $\gamma_{\rm eff}=
4/3$ and a 
suitable choice of the particle mass, the SPH resolution will thus always
be sufficient, leaving the gravitational softening as the limiting
factor. We
therefore use $\gamma_{\rm eff} = 4/3$ as our default value, but will
also explore the effect of varying the polytropic index.

Note that although our prescription for SF is designed to result in
a fixed KS law, regardless of the equation of state, this does not
mean that simulation predictions are independent of the equation of
state. The effective equation of state affects the gas distribution and
hence the SF history of ab initio simulations of the
formation of galaxies.

\section{Simulations}
\label{sec:sims}

\begin{table*}
\begin{center}
\caption{Simulations parameters:
  initial disc gas fraction, $f_{\rm g}$;
  total number of particles, $N_{\rm tot}$;
  total number of gas particles in the disc, $N_{\rm disc}$;
  mass of baryonic particles, $m_{\rm b}$;
  mass of dark matter particles, $m_{\rm DM}$;
  gravitational softening of baryonic particles, $\epsilon_{\rm b}$;
  gravitational softening of dark matter particles, $\epsilon_{\rm DM}$;
  effective polytropic index, $\gamma_{\rm eff}$;
  KS law exponent, $n$;
  gas surface density threshold for SF, $\Sigma_{\rm c}$;
  initial, central gas surface density, $\Sigma_0$;
  wind feedback included, (Wind). Values different from the fiducial ones are
  shown in bold.
\label{tbl:extra_params}}
\begin{tabular}{ccccccccccccc}

\hline
  Simulation & $f_{\rm g}$ & $N_{\rm tot}$ & $N_{\rm disc}$ & $m_{\rm
  b}$ & $m_{\rm DM}$ & $\epsilon_{\rm b}$ & $\epsilon_{\rm dm}$ & $\gamma_{\rm eff}$ & $n$ & $\Sigma_{\rm c}$ & $\Sigma_0$ & Wind \\
             &             &               &                &    $\Msolh$ &     $\Msolh$ &             $\pch$ &              $\pch$ &                    &     & $\Msolpcsq$ &  $\Msolpcsq$ & \\
\hline
  \textit{fid}       & 0.3 & 5,000,494 & 235,294 & $5.1\times 10^4$ & $2.4\times 10^5$ & 10 &  17 & $4/3$ & 1.4 & 7.3 & 228 & N \\
\hline
  \textit{f10}       & \textbf{0.1} & 5,000,494 & \textbf{78,431} & $5.1\times 10^4$ & $2.4\times 10^5$ & 10 &  17 & $4/3$ & 1.4 & 7.3 & \textbf{76} & N \\
\hline
  \textit{f90}       & \textbf{0.9} & 5,000,494 & \textbf{705,882} & $5.1\times 10^4$ & $2.4\times 10^5$ & 10 &  17 & $4/3$ & 1.4 & 7.3 & \textbf{684} & N \\
\hline
  \textit{gamma1}    & 0.3 & 5,000,494 & 235,294 & $5.1\times 10^4$ & $2.4\times 10^5$ & 10 &  17 &   \textbf{1} & 1.4 & 7.3 & 228 & N \\
\hline
  \textit{gamma5/3}  & 0.3 & 5,000,494 & 235,294 & $5.1\times 10^4$ & $2.4\times 10^5$ & 10 &  17 & \textbf{5/3} & 1.4 & 7.3 & 228 & N \\
\hline
  \textit{n1.7}      & 0.3 & 5,000,494 & 235,294 & $5.1\times 10^4$ & $2.4\times 10^5$ & 10 &  17 & $4/3$ & \textbf{1.7} & 7.3 & 228 & N \\
\hline
  \textit{sigma2.3}  & 0.3 & 5,000,494 & 235,294 & $5.1\times 10^4$ & $2.4\times 10^5$ & 10 &  17 & $4/3$ & 1.4 & \textbf{2.3} & 228 & N  \\
\hline
  \textit{sigma23}   & 0.3 & 5,000,494 & 235,294 & $5.1\times 10^4$ & $2.4\times 10^5$ & 10 &  17 & $4/3$ & 1.4 & \textbf{23} & 228 & N  \\
\hline
  \textit{wind}      & 0.3 & 5,000,494 & 235,294 & $5.1\times 10^4$ & $2.4\times 10^5$ & 10 &  17 & $4/3$ & 1.4 & 7.3 &  228 &\textbf{Y} \\
\hline
  \textit{lowres8}   & 0.3 & \textbf{625,059} & \textbf{29,411} & $\mathbf{4.1\times 10^5}$ & $\mathbf{1.9\times 10^6}$ & \textbf{20} &  \textbf{34} & $4/3$ & 1.4 & 7.3 & 228 & N \\
\hline
  \textit{lowres64}  & 0.3 &  \textbf{78,131} &  \textbf{3,676} & $\mathbf{3.3\times 10^6}$ & $\mathbf{1.5\times 10^7}$ & \textbf{40} &  \textbf{68} & $4/3$ & 1.4 & 7.3 & 228 & N \\
\hline
  \textit{lowres512} & 0.3 &   \textbf{9,765} &   \textbf{459} & $\mathbf{2.6\times 10^7}$ & $\mathbf{1.2\times 10^8}$ & \textbf{80} & \textbf{136} & $4/3$ & 1.4 & 7.3 & 228 & N \\
\hline

\end{tabular}
\end{center}
\end{table*}

We ran a suite of 12 numerical simulations of an isolated disc galaxy,
varying physical and numerical parameters to test the theoretical
analysis of the relation between the KS and Schmidt laws presented in
\S\ref{sec:analytics}, as well as its implementation as a recipe for
SF as described in \S\ref{sec:simrecipe}.
Before showing the results, we will describe the simulation setup and
the different runs we carried out.

\subsection{Code and initial conditions}

We used a modified version of the code \textsc{Gadget}
\citep{Springel2001,Springel2005}, which is a $N$-body
TreePM/SPH code. 

We implemented the prescription for SF described
in section~\ref{sec:simrecipe}. Unless stated otherwise, our models
used a SF threshold of $7.3~\Msolpcsq$, the observed KS law
(eq.~[\ref{eq:KS}]), a polytropic equation of state with index 
$\gamma_{\rm eff}=4/3$ normalised to $T/\mu=10^4~\K/0.59$ at the
threshold\footnote{We use $\mu=0.59$ because in
  our simulations, which include a UV background, the gas is still
  mostly ionized at the SF threshold because self-shielding is
  ignored. Our adopted equation of state implies pressure $P_{\rm
    tot}/k = 2.3\times 10^3~\cm^{-3}\,\K$ at the SF threshold.   
The value $\Sigma_{\rm c} = 7.3~\Msolpcsq$ corresponds to $n_{\rm H} =
0.1~\cm^{-3}$ for this equation 
of state and for our fiducial $f_{\rm g}=0.3$.}, and a
gas fraction $f_{\rm g}=0.3$ which is the initial disc gas fraction of our
fiducial model. 

A new module for feedback from star
formation\footnote{We use kinetic
  feedback, similar to 
that of \cite{Springel&Hernquist2003}, except that in our case wind
particles are selected locally and are not decoupled
hydrodynamically.}, which will be described elsewhere (Dalla Vecchia
et al., in  
preparation), was also implemented,  
but except for one of our runs, feedback was not included in the
simulations presented here. Radiative
cooling and heating was included using tables for hydrogen and helium,
assuming ionization equilibrium in the presence of the
\cite{Haardt&Madau2001} model for the $z=0$ UV background radiation
from quasars and galaxies. The cooling tables were generated using the
publicly available package \textsc{cloudy} \citep{Ferland2000}.

\begin{figure*}
\includegraphics[width=10.25pc]{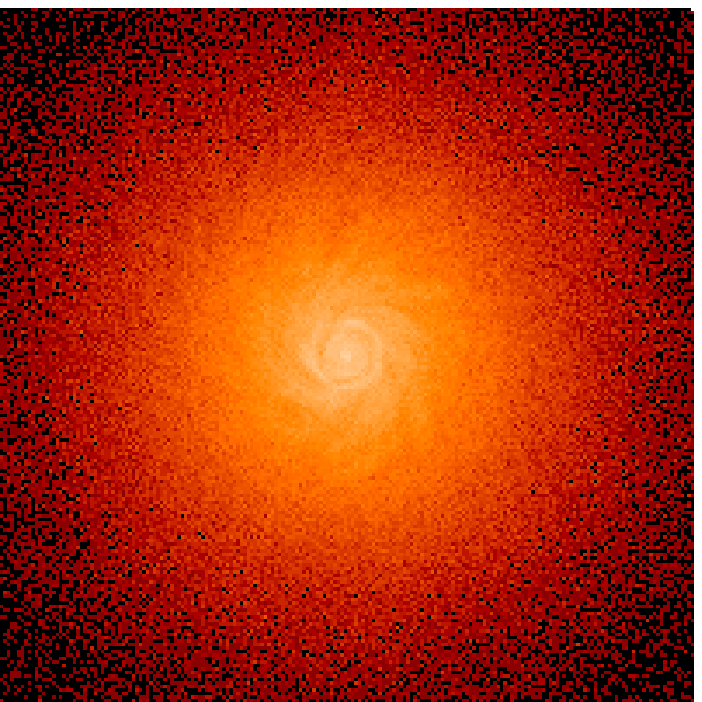}%
\includegraphics[width=10.25pc]{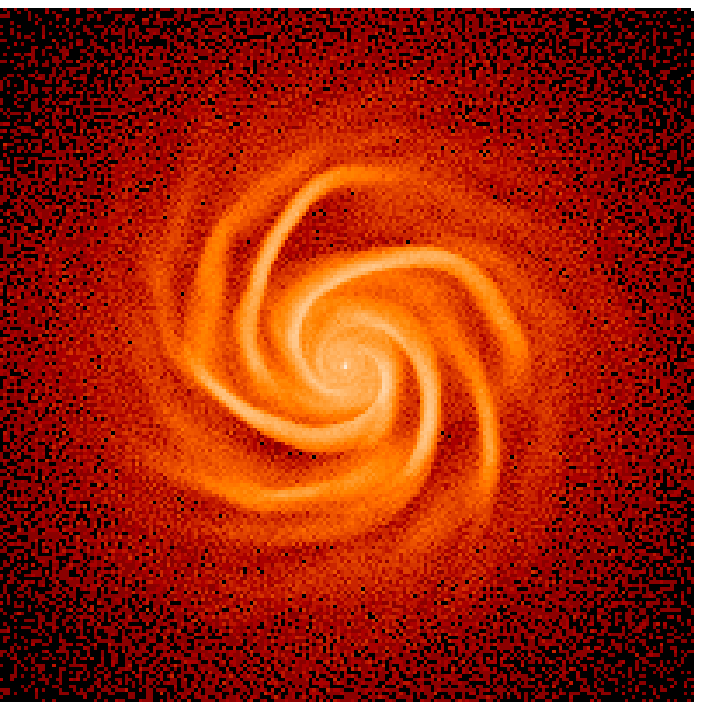}%
\includegraphics[width=10.25pc]{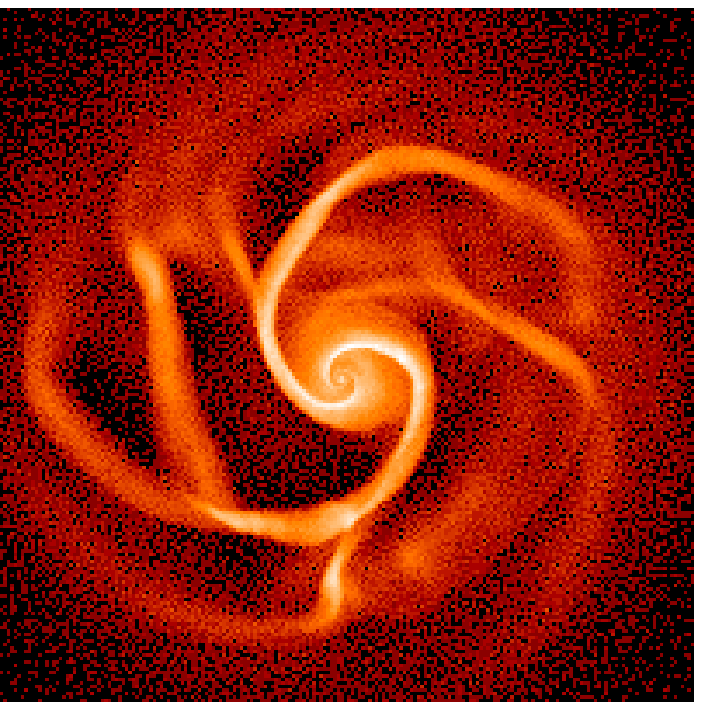}
\caption{Face-on projections of the disc gas distributions
  for model \textit{fid} at times, from left to right,
  $t=[25,125,250]~\Myr$. All images are $20~\kpch$ on a side
  and use the same logarithmic colour scale. The stellar disc, the
  bulge and the halo are not shown. 
\label{fig:sims_evol}}
\end{figure*}

The initial conditions were generated with a code kindly provided to
us by Volker Springel. A detailed description of the algorithm is given in
\cite{Springeletal2005}. Here we provide only a brief summary.

The model consists of a dark matter halo, a stellar bulge, and an
exponential disc containing both stars and gas. The total mass is
$10^{12}~h^{-1}\,\Msun$  (we use $h=0.73$) and the circular velocity
at the virial radius 
is $163~\kms$. The dark matter halo and the stellar bulge both
follow \cite{Hernquist1990} profiles. The Hernquist profile is scaled
to match the inner density profile of a \cite{NFW} profile with
concentration parameter $c=9$ and a mass within the virial radius
equal to the total mass of the Hernquist profile. The halo has a
dimensionless spin parameter $\lambda = 0.033$. The radial disc
scale length is $2.47~\kpch$ and was computed by relating it to the disc
angular momentum following \cite{MoMaoWhite1998,Springeletal2005}. The
disc contains 4~percent of both the total angular momentum and the
total mass. 
The bulge contains 1.4~percent of the
total mass and has a scale length one tenth of that of the disc. 
The bulge has no net rotation, while the dark matter halo and disc have
the same specific angular momentum. The vertical distribution of the
stellar disc follows the profile 
of an isothermal sheet with a constant scale height set to 10~percent of
the radial disc scale length (i.e.\ $0.247~\kpch$). The vertical gas
distribution is set up in hydrostatic equilibrium using an iterative
procedure. 

We set the mass of baryonic particles to
\begin{equation}
m_{\rm b} = {M_{\rm J}(\rho_{\rm g,c}) \over N_{\rm resol} N_{\rm ngb}},
\label{eq:m_b}
\end{equation}
where $M_{\rm J}(\rho_{\rm g,c})$ is the Jeans mass at the SF threshold for our
fiducial model (recall that 
the Jeans mass is independent of the density for $\gamma_{\rm eff}=4/3$), 
$N_{\rm ngb}=48$ is the number of particles within the SPH
kernel and $N_{\rm resol}$ is the factor by which the Jeans mass
exceeds the kernel mass. This choice of mass resolution implies
\begin{eqnarray}
h &=& \left ({3 N_{\rm ngb} m_{\rm b} \over 4\pi \rho_{\rm g}}\right )^{1/3} \\
&=& \left ({1 \over 8 N_{\rm resol}}\right )^{1/3} L_{\rm J}(\rho_{\rm g,c}) \left
(\rho_{\rm g} \over \rho_{\rm g,c}\right )^{2/3-\gamma_{\rm eff}/2}
\end{eqnarray}
where $h$ is the SPH kernel and we made use of equations
(\ref{eq:L_J}) and (\ref{eq:m_b}) and we assumed $f_{\rm g}$ to be
constant. For our fiducial value of
$\gamma_{\rm eff}$ this becomes $h/L_{\rm J} = (1/8N_{\rm resol})^{1/3}$,
independent of the density. We chose $N_{\rm resol}=6$ which
guarantees that we resolve the Jeans mass and (in terms of the SPH
smoothing) the Jeans length throughout the
disc, easily satisfying the resolution criterion of
\cite{Bate&Burkert1997}. Thus, our baryonic
particles have mass $m_{\rm b} = 
5.1\times 10^4~h^{-1}\,\Msun$. The dark matter
particle mass was taken to be higher by a factor
$(\Omega_{\rm m}-\Omega_{\rm b})/\Omega_{\rm b}\approx 4.6$. For most
of our runs the 
total number of particles in the simulation is 5,000,494, of which
235,294 are gas particles in the disc. 

The gravitational softening length was set to $\epsilon_{\rm b}=10~\pch$ for
 the baryons and to $(m_{\rm 
 dm}/m_{\rm b})^{1/3}\epsilon_{\rm b}\approx 17~\pch$ for the dark
 matter. This is 
sufficiently small to resolve the Jeans length by at least two
softening lengths up to gas surface densities of $\sim
10^{4.5-5}~\Msolpcsq$, which greatly exceeds our central surface
density of $\Sigma_0\approx 2.3 \times 10^2~\Msolpcsq$. Note, however,
that because our gravitational softening can be smaller than 
the SPH kernel, spurious fragmentation is possible in regions where
the local Jeans length is smaller than the SPH kernel
\cite[e.g.][]{Bate&Burkert1997}, as may happen for polytropic indices smaller
than $4/3$. 

\subsection{Runs}

Table~\ref{tbl:extra_params} lists all the simulations we have carried
out and gives the values of all the parameters that are varied. Our
fiducial simulation is labelled \textit{fid}. 
Models \textit{f10} and \textit{f90} use initial disc gas fractions of
10 and 90~percent, respectively, compared with
30~percent for our fiducial model. The number of particles used in
these runs is
scaled accordingly, so that the particle mass remains constant.
Models \textit{gamma1} and
\textit{gamma5/3} use effective polytropic indices $\gamma_{\rm
  eff}=1$ and $5/3$ (i.e.\ isothermal and adiabatic), respectively,
rather 
than our default value of $4/3$. Run \textit{n1.7} uses $n=1.7$ as
the power-law index of the KS law instead of our default value of
1.4. Models \textit{sigma2.3} and \textit{sigma23} have threshold
volume densities that are an order of magnitude lower and higher,
respectively, than our default value of $n_{\rm
  H,c}=0.1~\cm^{-3}$. Because we keep $T/\mu$ at the threshold fixed, this
yields critical surface densities that differ by $\sqrt{10}$ from the
fiducial value (see eq.~\ref{eq:SigmaJ_nH}). Model \textit{wind}
includes galactic winds, whereas the fiducial model does not. Our
recipe for generating galactic winds will be  
described elsewhere, but we note here that it results in much stronger
perturbations to the galaxy than the widely used model of
\cite{Springel&Hernquist2003}. Finally, runs \textit{lowres8}, 
\textit{lowres64}, and 
\textit{lowres512} use particle masses that are greater than those
used in our
fiducial model by factors of 8, 64 and 512, respectively. The spatial
resolution (i.e.\ the gravitational softening length) is increased by
factors of 2, 4 and 8, respectively.

\begin{figure*}
\includegraphics[width=10.25pc]{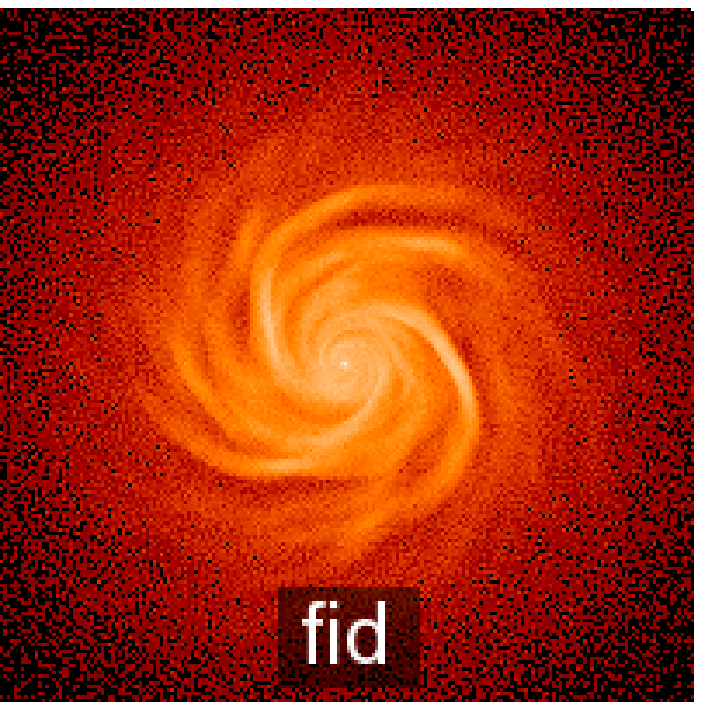}%
\includegraphics[width=10.25pc]{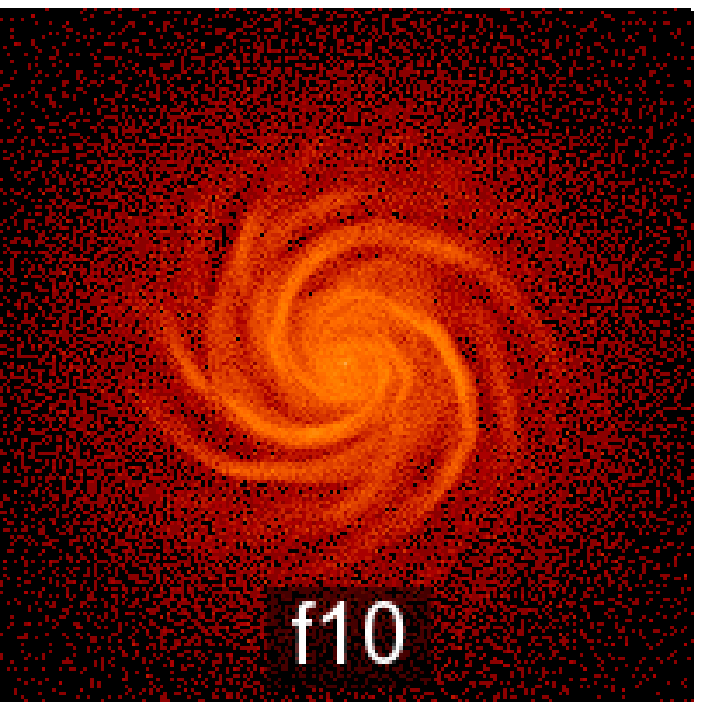}%
\includegraphics[width=10.25pc]{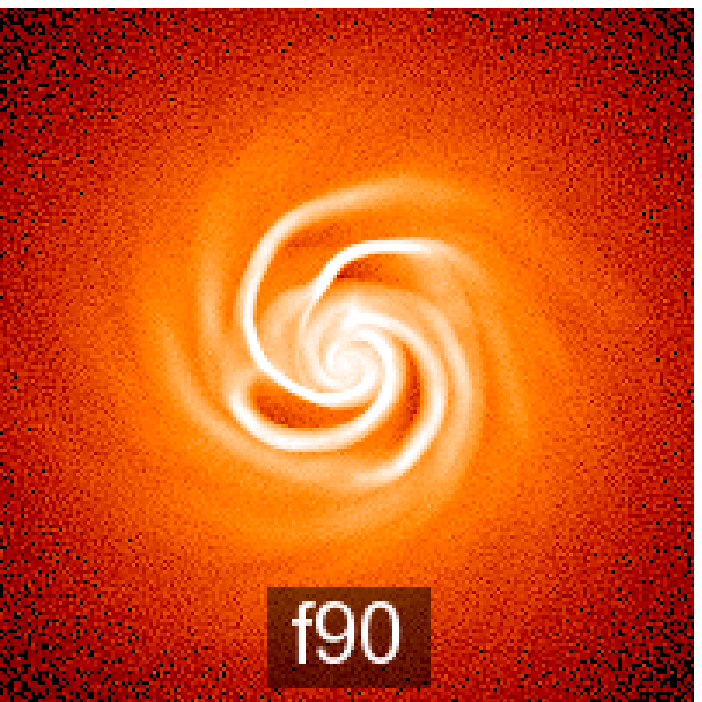}%

\includegraphics[width=10.25pc]{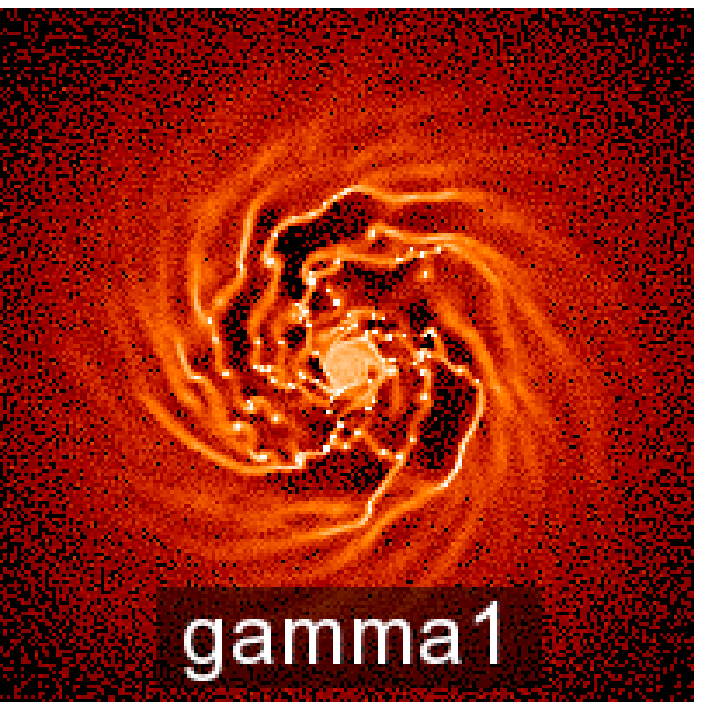}%
\includegraphics[width=10.25pc]{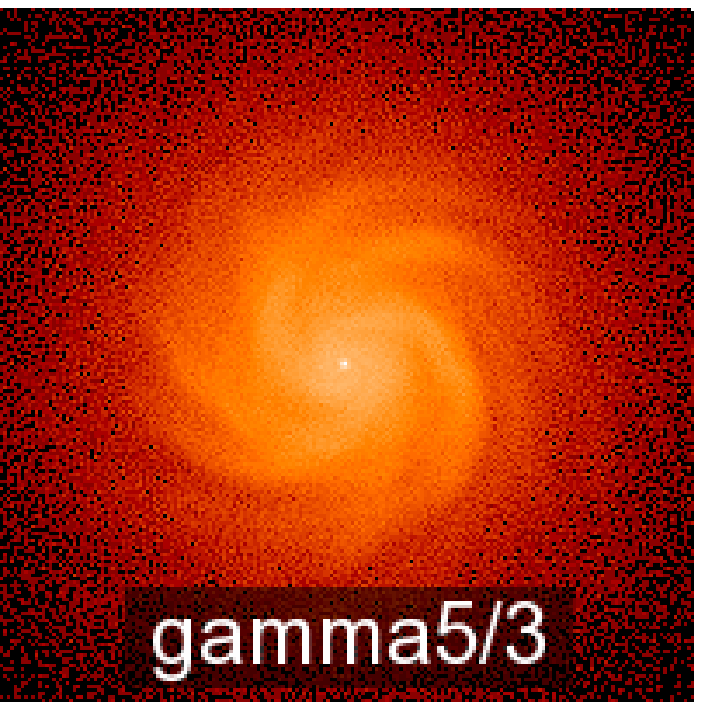}%
\includegraphics[width=10.25pc]{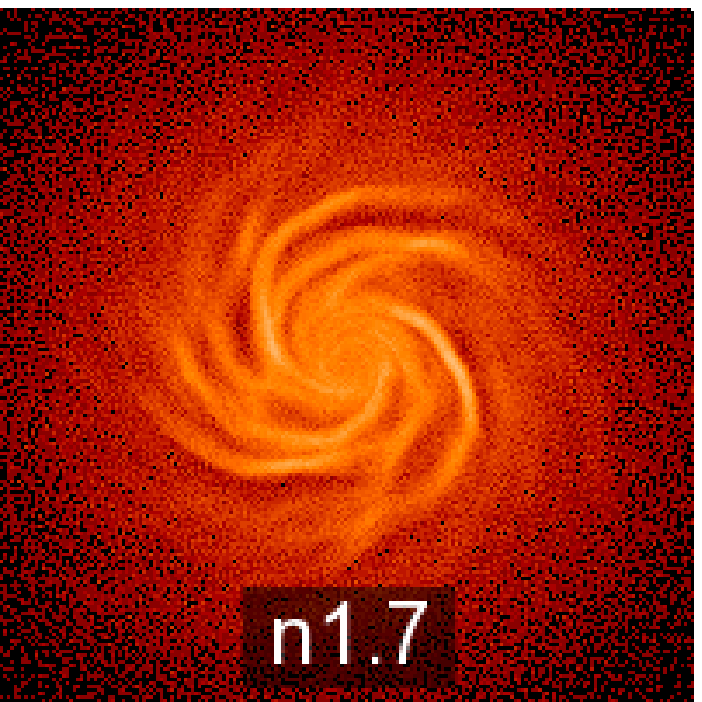}%

\includegraphics[width=10.25pc]{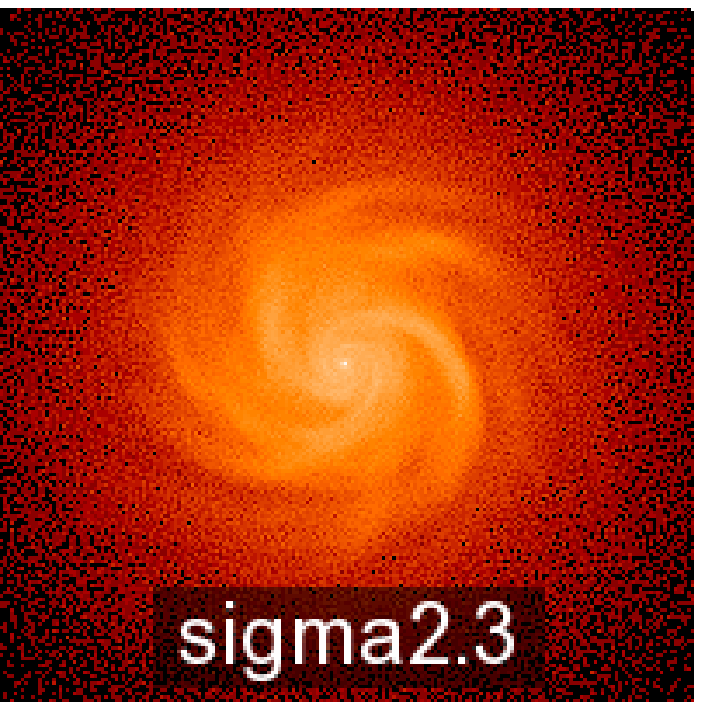}%
\includegraphics[width=10.25pc]{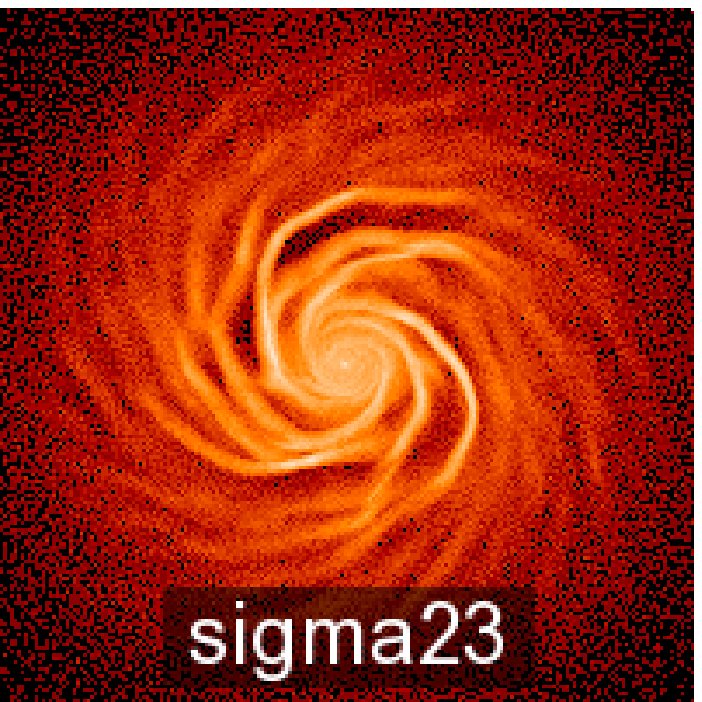}%
\includegraphics[width=10.25pc]{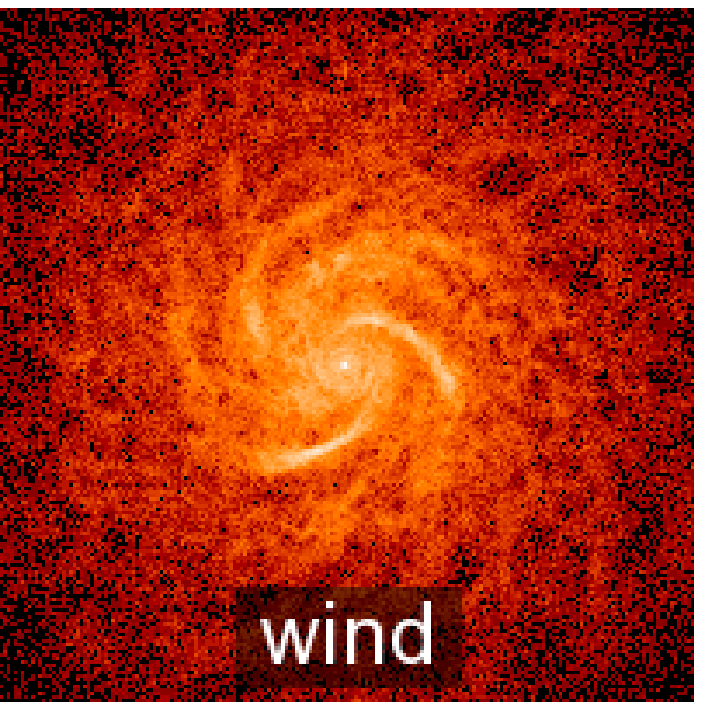}%

\includegraphics[width=10.25pc]{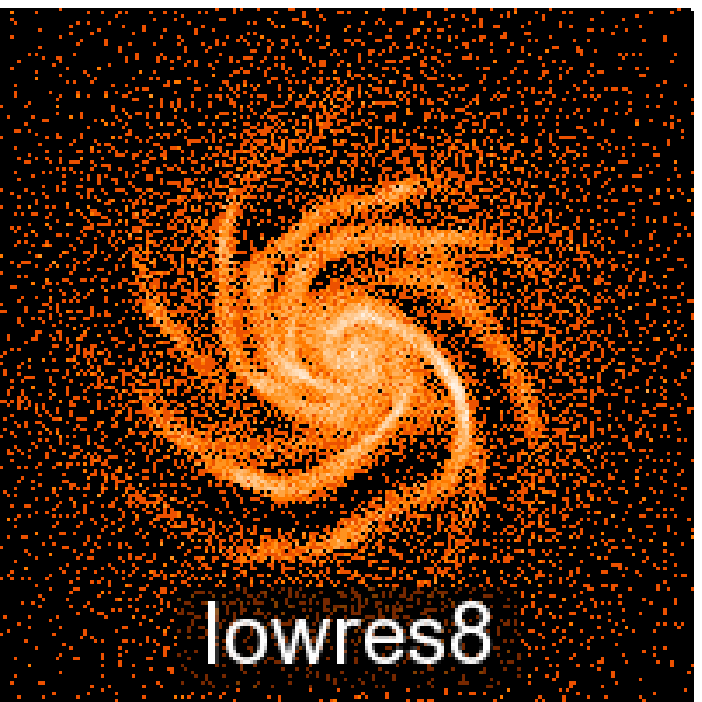}%
\includegraphics[width=10.25pc]{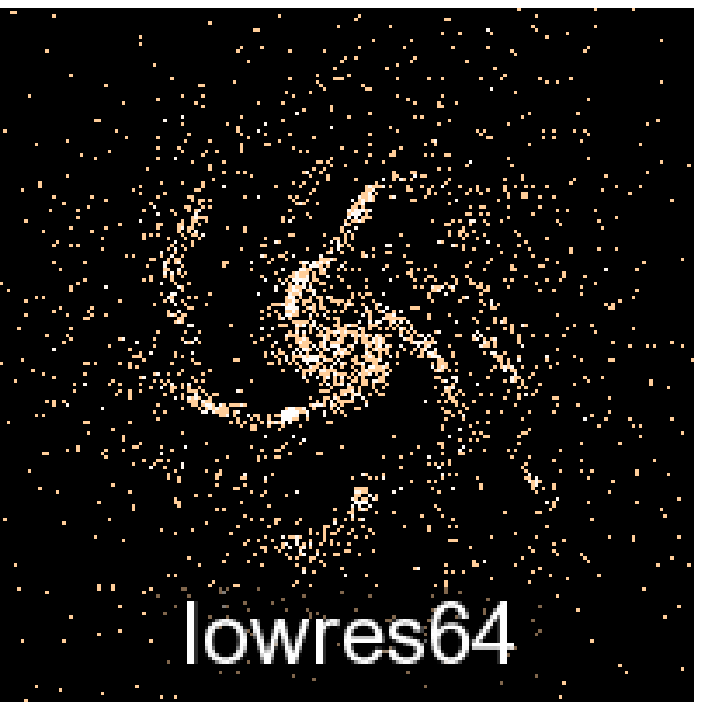}%
\includegraphics[width=10.25pc]{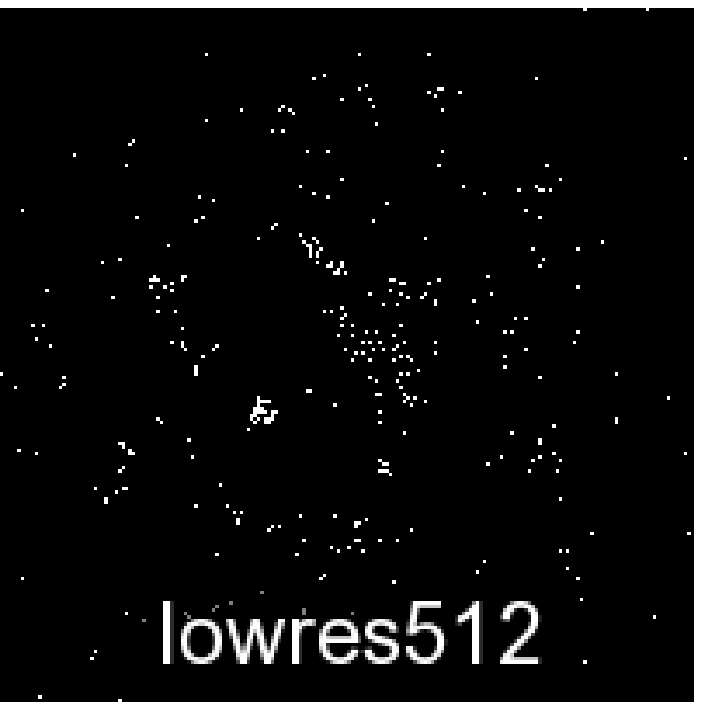}
\caption{Face-on projections of the disc gas surface densities
  for all models listed in Table~\protect\ref{tbl:extra_params}. All
  images are $20~\kpch$ on a side,  
  use the same logarithmic colour scale, and correspond to time
  $t=10^2$~Myr. The stellar disc, bulge and the halo are not
  shown. Each panel is labelled by the model ID, as used in
  Table~\protect\ref{tbl:extra_params}.} 
\label{fig:sims_100myr}
\end{figure*}

\begin{figure*}
\includegraphics[width=38pc]{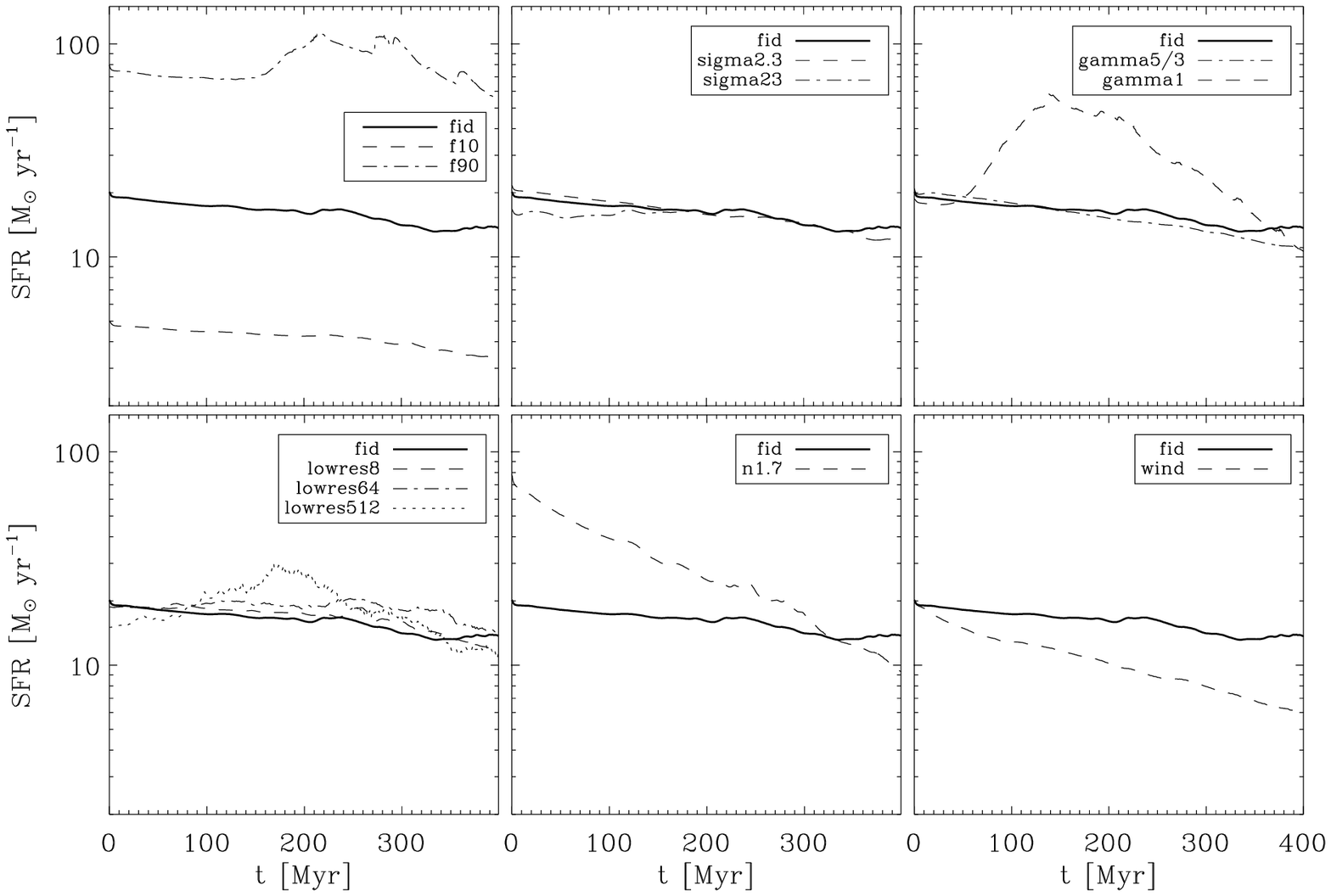}
\caption{Comparison of the SF histories of the different
simulations. The thick solid line in each plot shows the SF
history of our fiducial model, \textit{fid}. The other curves are
labelled in the legends of the individual panels.}
\label{fig:sfh_panel}
\end{figure*}

\subsection{Results}

\subsubsection{Global properties}

Figure~\ref{fig:sims_evol} shows how our fiducial model evolves. From left to
right, the panels show face-on projections of the disc gas
distribution at times 25, 125, and 250~Myr 
after the start of the run. Spiral arms develop rapidly, even though
the discs were initially axisymmetric. By time $T=250$~Myr, the spiral
arms are starting to fragment.

Figure~\ref{fig:sims_100myr} compares the face-on projections of the disc
gas surface density for all runs.  Because our galaxies have a fixed
disc mass, a higher gas fraction implies higher gas surface
densities, which results in sharper spiral arms. Increasing the
polytropic index makes the galaxies more diffuse, which is not
surprising given that 
both the Jeans length (\ref{eq:L_J}) and mass (\ref{eq:M_J}) are
increasing functions of $\gamma_{\rm eff}$. On the
other hand, model
\textit{gamma1}, which uses an isothermal effective equation of state,
becomes unstable and fragments in small clumps. The same behaviour is visible
in the models of \cite{Li2005} and \cite{Springeletal2005}. 
As we will demonstrate below, this simulation reproduces the desired local
KS law, despite the strong instability. Increasing the slope of
the KS law does not have much effect on the morphology of the
galaxy. Including galactic winds 
makes the spiral arms more diffuse and results in the generation of
low density bubbles in the outer disc. Increasing (decreasing) the
SF 
threshold results in sharper (more diffuse) spiral arms. Again, this
can be understood in terms of the Jeans scales, which decrease with
the gas density. While model \textit{lowres8} still looks reasonably
smooth, spurious fragmentation is visible in runs \textit{lowres64}
and particularly in \textit{lowres512}.

We show in Figure~\ref{fig:sfh_panel} a comparison of the SF
histories integrated over the entire galaxy from $t=0$
till $t=400$~Myr, which 
is when we stopped the runs. The SFR of the fiducial simulation
declines gently due to gas consumption. The same behaviour is seen in
most of the other runs, with the exception of the models that we
showed to be unstable in Fig.~\ref{fig:sims_100myr}, 
\textit{gamma1} and \textit{lowres512}, as well as \textit{f90} which becomes
unstable shortly after the snapshot shown in
Fig.~\ref{fig:sims_100myr}. For these 
unstable models the SFR increases due to fragmentation, although gas
consumption eventually forces it to decline. 

Models \textit{f10} and \textit{f90} have initial gas fractions that
differ by a factor of 3 from model \textit{fid}. Since the total disc mass
is fixed, this means that the gas surface densities differ by the same
factor. Neglecting the change in the total gas mass above the SF
threshold, which is justified since the SFR is dominated by dense gas,
we would expect the SFRs to differ initially by about a 
factor $3^{1.4}$. This corresponds to 0.67~dex and agrees nearly perfectly with
the simulation results. 

The SF history of model \textit{sigma2.3}, which has a critical surface
density smaller than the fiducial model by 0.5~dex, is nearly
indistinguishable from that of the fiducial run. This is because
surface densities below the fiducial threshold contribute very little
to the total SFR. As expected, model \textit{sigma23} is initially
below the fiducial one because of the higher SF threshold. As the gas
collapses to higher densities, so does the total gas mass above the
threshold surface density. The SFR therefore increases slightly,
temporarily 
surpassing that of model \textit{fid} because it is forming stars in gas at
higher surface densities, until gas consumption forces it to decline.

Run \textit{gamma5/3}, which has a stiffer equation of state,
is very similar to model \textit{fid}. Note, however,
that ab initio models (i.e.\ simulations which model the formation of galaxies)
could still show significant differences between the SF histories of
simulations using different equations of 
state. This would, however, not mean 
that the predicted KS laws are different, which we shall show below is not the
case. It would merely indicate that the equation of state affects the
gas distribution. Indeed, run \textit{gamma1}, which has a softer
equation of state, has a much higher SFR because the gas distribution
changed as a result of fragmentation. But, as we shall show
below, the local KS law still holds. 

Decreasing the mass (spatial) resolution by a factor 8 (2) does not
result in a significantly different SF history, indicating that our fiducial
simulation has converged. Decreasing the resolution further does,
however, lead to differences because the disc becomes unstable. Again,
we shall see that this does not imply that the KS laws differ.

Increasing the slope of the KS law naturally leads to a higher SFR,
although it eventually drops below the fiducial rate due to the fast
gas consumption. Finally, including feedback from SF in
the form of winds reduces the SFR. It takes some time for the
difference to develop because the initial conditions were the same,
i.e., the starting point is not affected by the winds.

\subsubsection{Local properties}

\begin{figure*}
\includegraphics[width=38pc]{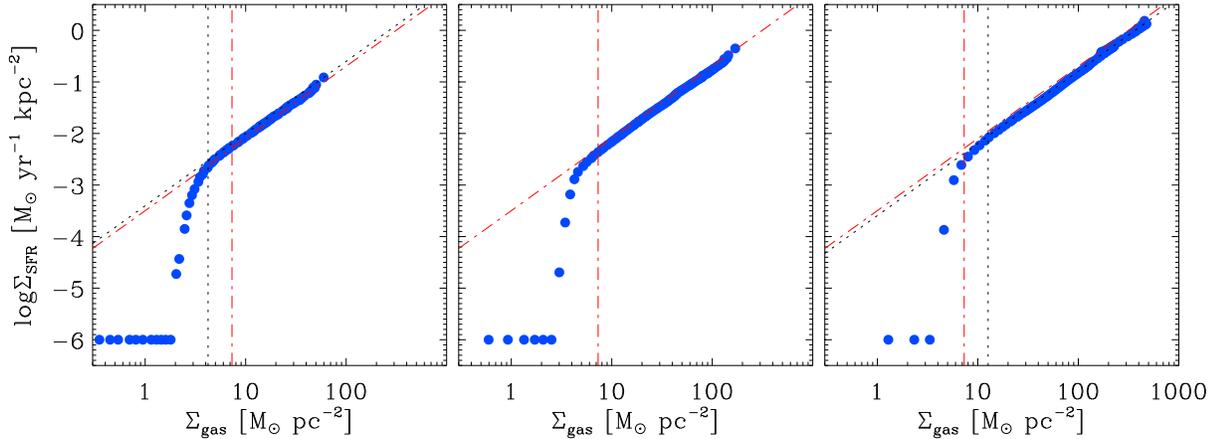}
\caption{KS law for simulations \textit{f10} (left
panel), \textit{fid} (middle panel) and
\textit{f90} (right panel) at $t=100~\Myr$. Filled circles indicate
SFR per unit area as a function of the gas surface density, both
averaged azimuthally in radial bins containing a a fixed gas mass. The
dot-dashed lines show the input KS law and threshold surface density
for the fiducial simulation. The dotted lines show the expected
scaling with the gas fraction (see the text for details). Data points with
$\dot{\Sigma}_\ast < 10^{-6}~\Msolyrkpcsq$ are set equal to this value
for clarity. The agreement between the input SF law and the simulation
results is clearly excellent.}
\label{fig:kenn_fg}
\end{figure*}

\begin{figure*}
\includegraphics[width=38pc]{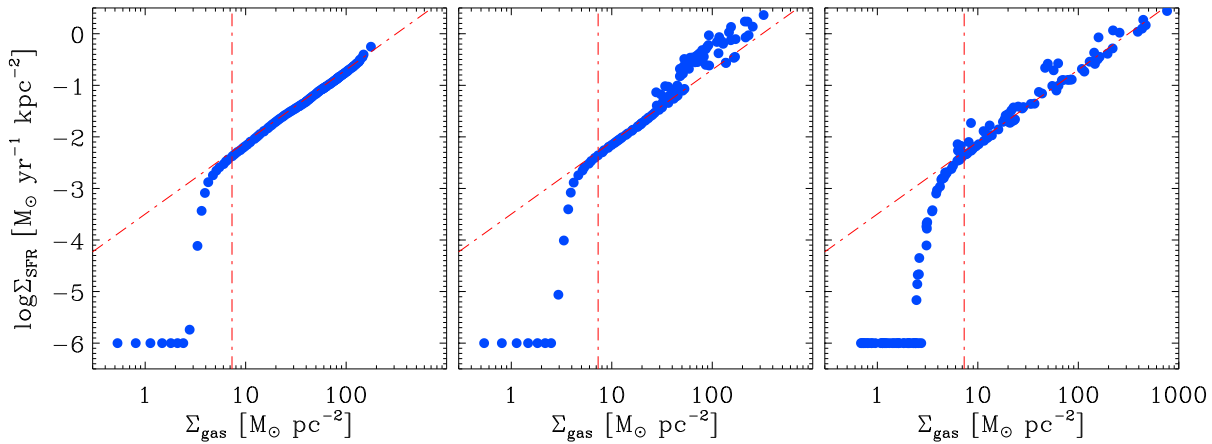}
\caption{As Fig.~\protect\ref{fig:kenn_fg} but for simulations
  \textit{gamma5/3} (left 
panel), \textit{gamma1} (middle panel)
and \textit{gamma1} with cylindrical binning (right panel). The local
  KS law reproduces the input law everywhere, but for unstable regions
  of the disc this is only visible if we use local bins rather than
  azimuthal averages.}
\label{fig:kenn_eos}
\end{figure*}

We now come to the most relevant results from the simulations, the
SF threshold and the local KS law. Unless noted otherwise,
we tested the KS law by  
projecting the gas density and SFR along the minor
axis, and measuring their values in radial, concentric bins containing
a fixed gas mass (corresponding to 2000 particles for run
\textit{fid}). For 
comparison, we explored two other ways of binning the data:
radial bins spanning an azimuthal angle $\Delta\phi \ll 2\pi$ and 
randomly distributed cylinders containing a fixed number of
particles. All three methods 
give nearly identical results if the disc
is stable. All plots in this section are for $t=100$~Myr and can
therefore be compared directly to Figure~\ref{fig:sims_100myr}.

The data points in Figure~\ref{fig:kenn_fg} show the SFR per unit area as a
function of the gas surface density, both averaged azimuthally, for 
simulations \textit{f10}, \textit{fid} and \textit{f90}. In this
figure and also in the remaining ones, all bins for
which $\dot{\Sigma}_\ast < 10^{-6}~\Msolyrkpcsq$ have been set equal
to this value for clarity. The red
dot-dashed lines are not fits to the data points, but show the input
KS law and SF threshold.

The agreement between the simulation results and the input KS law is
nearly perfect, both in terms of its normalisation and slope. In
fact, the agreement is better than could have been expected, given
that the analytic formulas derived in \S\ref{sec:analytics_KSlaws}
are no more accurate than the arbitrariness in the definition of the
Jeans length, roughly a factor $\sqrt{\pi}$. However, from
equations (\ref{eq:SigmaJ_nH}) and (\ref{eq:Schmidt}), we can see that
this would only change the normalisation of the Schmidt law by a
factor $\pi^{(n-1)/2}$, which is 0.1~dex for $n=1.4$, and that the
slope would not be affected at all. Furthermore, S04 showed that for our
definition of the Jeans length, (\ref{eq:SigmaJ_nH}) agrees with the exact
solution for an isothermal, gaseous disc 
to within 2~percent. 

The SFR clearly declines relatively sharply below a surface density
threshold that is in good agreement with the input value. The fact
that the agreement is less good for the threshold than for the KS law
is not unexpected. For a fixed volume density, the corresponding
surface density is proportional to the Jeans length. Therefore,
discrepancies of the order of $\sqrt{\pi}$ (0.6 dex) are to be
expected, but the difference is actually somewhat smaller than that,
Note that the sharpness of
the radial cut off in the SFR is sensitive to the way in which the
simulation results are binned. Because the galaxies are not axisymmetric,
azimuthal averaging smooths out the threshold, even if it is sharp
locally.

All three simulations use the same values for the threshold volume
density $\rho_{\rm g,c}$ and for the normalization of the SF law. That is,
we always used the fiducial value $f_{\rm g}=0.3$ for the calculation of
$\rho_{\rm g,c}$ from the input surface density threshold using equation
(\ref{eq:SigmaJ_nH}), and also in the implemented KS law, equation
(\ref{eq:dotmstar}). We neglected the weak dependence of the SF law on
the gas fraction because we only know the \emph{initial disc} gas
fraction. The gas fraction appearing in the equations depends
on both radius (due to the non-negligible contribution of the bulge near the
center and the dark halo in the outer parts) and time (due to gas
consumption).

The dotted lines in the left and right panels of
Fig.~\ref{fig:kenn_fg} show the expected SF threshold and KS law
assuming that the gas fraction appearing in the equations is identical
to the initial disc gas fraction. As for the uncertainty due to the
arbitrariness in the definition of the Jeans length, the dependence on
the gas fraction is only significant for the critical surface
density and even here it is weak. While run \textit{f10} reproduces
the expected threshold better if the gas fraction is taken into
account, the opposite is true for run \textit{f90}. It appears that the
initial disc gas fraction does change the threshold surface density, but
the effect is weaker than the naive scaling indicated by the dotted
lines. 

Changing the effective equation of state of the star forming gas has
the largest impact on the appearance of the gas disc
(Fig.~\ref{fig:sims_100myr}). Figure~\ref{fig:kenn_eos} shows the
adiabatic (\textit{gamma5/3}, left panel) and isothermal
(\textit{gamma1}, middle panel) runs, which can also be compared with
the fiducial model shown in the middle panel of Fig.~\ref{fig:kenn_fg}.
The adiabatic and fiducial runs are nearly identical and both agree
extremely well with the input SF law. This shows that we have
succeeded in implementing local KS laws in a way that is independent
of the effective equation of state. Clearly, in this respect our
analytic theory describes the simulated galaxies very well. 

At first sight, the isothermal run (middle panel) appears to
overestimate the SFR by factors of a few for high surface
densities ($\Sigma_{\rm g}>20~\Msolpcsq$). The disagreement is limited to
the region of the disc that 
is unstable (Fig.~\ref{fig:sims_100myr}). However, what we are seeing
here is not a break-down of our theory, but rather an artifact caused by
azimuthal smoothing. In the unstable region the departures from
axisymmetry become very large and, at a fixed radius, the surface
density fluctuates strongly. Because the KS law is supra-linear ($n>
1$), azimuthal averaging will result in the inference of a steeper KS
law. 

We can remedy the binning problem by avoiding azimuthal smoothing. In
the right panel of Fig.~\ref{fig:kenn_eos} we show the results for
cylindrical binning. We use 
an equal number of bins as in the other panels, but center them on
randomly chosen gas particles. For each
bin we average $\dot{\Sigma}_\ast$ and $\Sigma_{\rm g}$ over a cylinder that
contains a fixed number of particles. We chose bins of 100 particles
as a compromise between the need to suppress noise (smaller for bigger
bins) and the need for small spatial bins (using $\gg 100$ particles
per bin leads to the same problem as we had for azimuthal binning). Clearly,
purely local binning, which is more appropriate but more noisy
than azimuthal binning, reveals that even this manifestly unstable
disc follows the local KS law exactly as predicted.

\begin{figure}
\center\includegraphics[width=16.2pc]{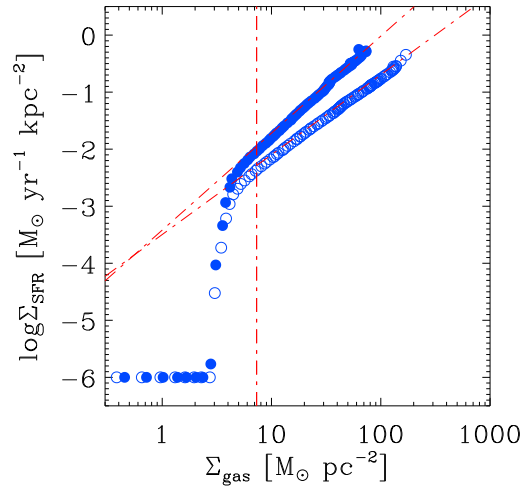}
\caption{As Fig.~\protect\ref{fig:kenn_fg} but for models
  \textit{n1.7} (filled circles) and \textit{fid} (open
  circles). The prescription for SF can match different input slopes
  of the KS law.}   
\label{fig:kenn_schm}
\end{figure}

Figure~\ref{fig:kenn_schm} demonstrates that the success of our model
is not confined to one particular form of the KS law. If we change the
slope of the input law from $n=1.4$ (open data points) to $n=1.7$
(filled data points), the simulation still reproduces the expected
relation. As before, the very small deviations at high surface
densities are caused by the decrease in the gas fraction due to gas
consumption. For model 
\textit{n1.7} this happens at a smaller surface density because it
has a higher SFR (for fixed $\Sigma_{\rm g}$). 

\begin{figure}
\center\includegraphics[width=16.2pc]{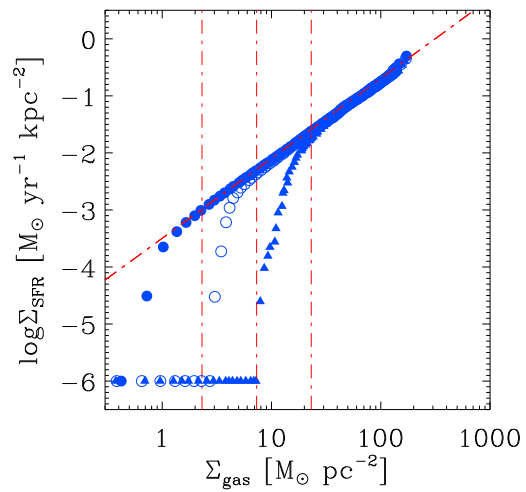}
\caption{As Fig.~\protect\ref{fig:kenn_fg} but for models
  \textit{sigma2.3} (filled 
circles), \textit{fid} (open circles) and \textit{sigma23} (filled
  triangles). The models match the desired SF thresholds.}
\label{fig:kenn_thr}
\end{figure}

Figure~\ref{fig:kenn_thr} shows the
effect of changing the surface density threshold. Comparing runs
\textit{sigma2.3} (filled circles), \textit{fid} (open circles) and
\textit{sigma23} (filled triangles), it is clear that $\Sigma_{\rm c}$
scales as predicted.

\begin{figure*}
\center\includegraphics[width=26.5pc]{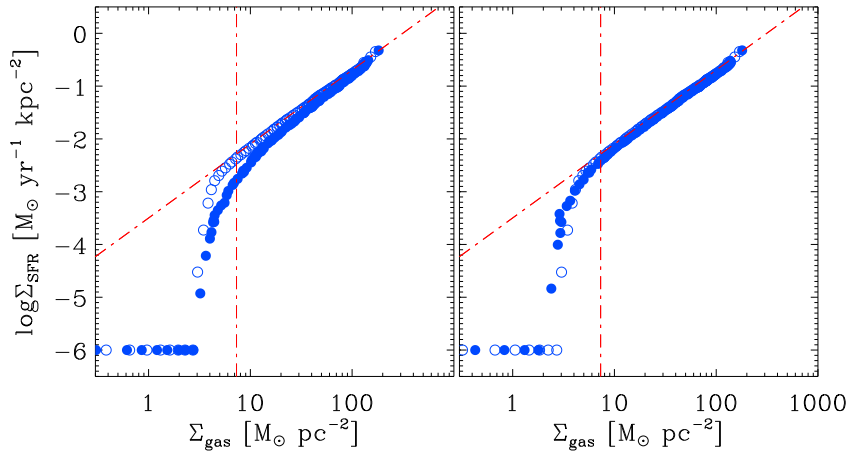}
\caption{As Fig.~\protect\ref{fig:kenn_fg} but for models \textit{wind} (filled
circles) and \textit{fid} (open circles). In the right hand panel
$\dot{\Sigma}_\ast$ and $\Sigma_{\rm g}$ include only gas within a scale
height of 1~kpc, while in the left panel we integrate to infinity, as
was in the other figures. 
The model including galactic
winds matches the input KS law very well, although it falls slightly
below the fiducial model for densities around the threshold
value. Excluding ejected gas (right panel) removes even this small
discrepancy.}
\label{fig:kenn_wind}
\end{figure*}

It may be argued that our model, which relies on the assumption that
galaxy discs are self-gravitating, would break down for galaxies that
are perturbed, although the results for the strongly unstable models
suggest otherwise. It is therefore interesting to see how the model
fares if we introduce a violent form of feedback from
SF. As can be seen from Figures~\ref{fig:sims_100myr} and
\ref{fig:sfh_panel}, our prescription 
for winds (which we will describe elsewhere) clearly has a dramatic
effect on both the appearance and the SF history of the
galaxy. Nevertheless, as is shown in the left panel of
Figure~\ref{fig:kenn_wind}, the 
simulation including winds (filled circles) still agrees with the
input KS law with very little 
scatter. For surface densities around the threshold, the predicted SFR
falls somewhat below the input KS law. The reason is that at these low
densities the wind energy is 
dissipated sufficiently slowly that a significant part of the gas surface
density is made up of ejected gas, which resides far above the disc
and has volume densities that are too low to partake in the SF. In the
right-hand panel of 
Figure~\ref{fig:kenn_wind} we again show models \textit{fid} and
\textit{wind}, but this time we integrate only out to a scale height
of 1~kpc. While this makes no difference for our fiducial model, it
brings model \textit{wind} in nearly perfect agreement with the input
KS law.

\begin{figure*}
\includegraphics[width=38pc]{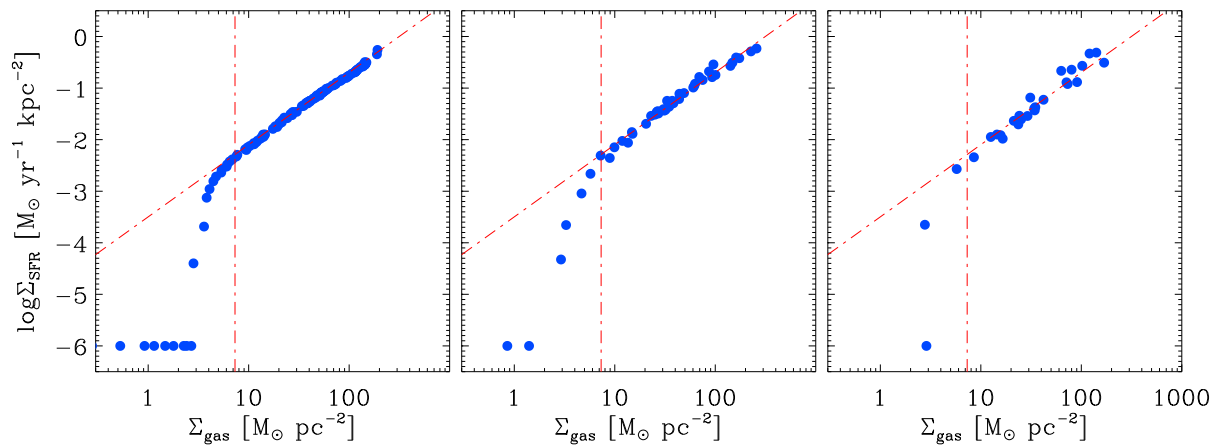}
\caption{As Fig.~\protect\ref{fig:kenn_fg} but for models
  \textit{lowres8} (left 
panel), \textit{lowres64} (middle panel)
and \textit{lowres512} (right panel). The simulations match the input
  SF law, even for very low resolutions.}
\label{fig:kenn_res}
\end{figure*}

Finally, we show the results of a convergence test in
Figure~\ref{fig:kenn_res}. 
Reducing the mass resolution by factors of 8 (\textit{lowres8}, left
panel), 64 (\textit{lowres64}, middle panel) or even 512 
(\textit{lowres512}, right panel) does not undermine the agreement
between the simulation results and the input SF law. This is quite
remarkable, given that the two lowest resolution runs suffer strongly
from spurious fragmentation (see Fig.~\ref{fig:sims_100myr}). Note
that the slight discrepancy for the lowest resolution model is most
likely due to azimuthal binning (we cannot verify this because there
are not enough particles for spatially resolved binning).

\section{Discussion and conclusions}

Star formation (SF) is poorly understood, but observations have
revealed the existence of useful laws to describe its behaviour
when averaged over scales that are large compared to individual star
clusters. SF is 
suppressed if the local gas surface density falls below $\Sigma_{\rm c}\sim
3$--$10~\Msolpcsq$, while for higher densities a Kennicutt-Schmidt (KS)
law, i.e.\ $\dot{\Sigma}_\ast\propto \Sigma_{\rm g}^n$, is a good
description of the data. When averaged over entire galaxies, $n=1.4$
is a good fit and the same law may also hold locally, although this is
less well established. If the local KS law is supra-linear ($n> 1$), then
the global law can only be identical if the total SF rate is dominated
by a small region, e.g.\ the nucleus. 

In S04 we provided a theoretical explanation for the existence of SF
thresholds, showing that the surface density threshold for the
formation of a cold gas phase agrees with the observationally
determined SF threshold and that this transition triggers
gravitational instability. S04 also provided a method for the
implementation of arbitrary surface density thresholds in numerical
simulations. Here, we have extended the work of S04 to KS
laws. Although we have not offered an 
explanation for the observed KS law, we have provided a theory that
relates various SF laws, that sheds new light on previous attempts
to account for the observed KS law, and that enables the
implementation of arbitrary SF laws into simulations of the formation
and evolution of galaxies.

We showed that the KS law is primarily a pressure law since, as is
well known, $\Sigma_{\rm g} \propto P_{\rm tot}^{1/2}$ for a
self-gravitating disc, although the relation depends also on the gas
fraction. The KS law is related to the Schmidt law, i.e.\
$\dot{\rho}_\ast \propto \rho_{\rm g}^{n_S}$, but the  relation depends on
the scale height and its dependence on the surface density. In
general, the relation between surface and volume densities depends on 
the effective equation of state 
of the multiphase ISM and also on both the gas fraction and the
fraction of the pressure that is thermal: $\Sigma_{\rm g} \propto
f_{\rm g}^{1/2}f_{\rm th}^{-1/2} T^{1/2} \rho_{\rm g}^{1/2}$. For a polytropic
effective equation of state, i.e.\ $P_{\rm tot}\propto
\rho_{\rm g}^{\gamma_{\rm eff}}$, 
the power-law indices of the Schmidt and KS laws are related via $n =
1+2(n_S-1)/\gamma_{\rm eff}$ if the gas fraction is independent of the
density. More generally, the various SF laws, including their
normalisations, can be related via the equations given in
section~\ref{sec:analytics}.

Unless the SF laws are linear (i.e.\ $n=1=n_S$), the slopes of the
Schmidt and KS 
laws will generally differ and the relation between the two will
depend on the effective equation of state. Theoretical models have often ignored
this difference and the dependence on the effective equation of
state. For example, simulations and semi-analytic models typically assume
a $n_S = 1.5$ Schmidt law motivated by the fact that it implies a gas
consumption time-scale proportional to the dynamical time
($\rho_{\rm g}/\dot{\rho}_\ast\propto \rho_{\rm g}^{-1/2}$) if, and this is
sometimes forgotten, the SF efficiency per dynamical time is
independent of the density. It is often mentioned that $n_S=1.5$ is
also consistent with the observed KS law, $n=1.4\pm 0.1$. However,
for $n_S=1.5$ to give $n=1.4$ we require $\gamma_{\rm eff}=2.5$ which
is extreme and much greater than assumed/predicted by any of the
simulations.  

As long as simulations of galaxies lack the physics and/or resolution
to model the 
multiphase ISM, in particular molecular cooling, radiative transfer, and
efficient feedback from stellar winds and supernovae, they will
require prescriptions for SF that can be calibrated to match observed
scaling laws. We argued that such simulations should use an
effective equation of state above the SF threshold, where the ISM is
predicted to be multiphase. A polytropic index of $\gamma_{\rm eff}=4/3$ is
particularly attractive from a numerical point of view since it
results in a constant Jeans mass (and, for the case of SPH
simulations, a constant ratio of the SPH kernel to the Jeans length),
while the 
Jeans length still decreases with density. This choice therefore
minimises resolution effects without directly suppressing collapse.

Because the observed SF laws involve surface
densities whereas simulations require volume densities or pressures,
our work is ideally suited for the implementation of arbitrary KS laws into
simulations of galaxies. In particular, equation
(\ref{eq:mstardot}) expresses the star formation rate per resolution
element in terms of its total total pressure and
can thus be used even if the polytropic index depends on the density
or if no effective equation of state is imposed.

Our equations enable us to predict
many of the predictions of such simulations regarding large-scale SF
laws, even though our theoretical framework does not explain why real galaxies
follow such laws. The fact that we can predict simulation predictions
once we know the sub-grid SF implementation, 
implies that simulations of galaxies that
do not try to actually simulate the multiphase ISM, can only provide
limited insight into the origin of the observed SF laws. 

We discussed how to 
implement arbitrary KS laws into simulations that do not attempt to
simulate the multiphase ISM and did so ourselves
for the smoothed particle hydrodynamics code \textsc{gadget}. 
We used this code to test our analytical relations
using high-resolution simulations of an isolated disc galaxy. We
found that the simulations follow the predicted relations nearly
perfectly. The surface density threshold and both the normalisation
and the slope of the KS law are reproduced with hardly any
scatter and without any tunable parameters. Moreover, this success is
not limited to the SF laws that happen to describe the real universe, 
we can reproduce arbitrary input laws, again without tuning any 
parameters. 

The agreement between theory
and numerical experiment remains excellent if we include violent
feedback from SF or use extremely low 
resolution. The only significant deviations that we found, occurred in
unstable galaxies in which the spiral arms have fragmented. However,
even these manifestly unstable galaxies turned 
out to follow the theoretical relations closely. The apparent
discrepancy between the input and output SF laws turned out to be an
artifact of azimuthal smoothing, which becomes inappropriate once
the galaxy becomes fragmented.

Finally, we would like to stress that to better constrain the models,
it is extremely important to get
better observational constraints on 
the \emph{local} SF threshold and the \emph{local} KS law for a
representative sample of galaxies. Even
azimuthal smoothing 
should be avoided if at all possible, as it greatly confuses the
comparison between theory and observations.

\section*{Acknowledgments}

We are very grateful to Volker Springel for allowing us to use GADGET
and his initial conditions code for the simulations presented here, as well
as for useful discussions, help with his codes, and a careful reading
of the manuscript. We also gratefully acknowledge discussions 
with the other members of the OWLS and Virgo collaborations. We would
also like to thank the referee, Ralf Klessen, and Andrey Kravtsov for
useful comments that helped improve the manuscript. The
simulations presented here were run on the Cosmology
Machine at the Institute for Computational Cosmology in Durham as part
of the Virgo Consortium research programme and on Stella, the
LOFAR BlueGene/L system in Groningen. This work
was supported by Marie Curie Excellence Grant MEXT-CT-2004-014112.


\begin{thebibliography}{MBNB03}

\bibitem[\protect\citeauthoryear{Auld et al.}{2006}]{Auld2006} 
Auld R., de Blok W.~J.~G., Bell E., Davies J.~I., 2006, MNRAS, 366, 1475 

\bibitem[\protect\citeauthoryear{Bate \& 
Burkert}{1997}]{Bate&Burkert1997} Bate M.~R., Burkert A., 1997, MNRAS, 
288, 1060 

\bibitem[\protect\citeauthoryear{Blitz \& 
Rosolowsky}{2006}]{Blitz&Rosolowsky2006} Blitz L., Rosolowsky E., 2006, ApJ, 
650, 933 

\bibitem[\protect\citeauthoryear{Boissier et 
al.}{2006}]{Boissier2006} Boissier S., et al., 2006, astro, 
arXiv:astro-ph/0609071 

\bibitem[\protect\citeauthoryear{Booth, Theuns, \& 
Okamoto}{2007}]{Booth2007} Booth C.~M., Theuns T., Okamoto T., 
2007, MNRAS, 376, 1588 

\bibitem[\protect\citeauthoryear{Cen \& 
Ostriker}{1992}]{Cen&Ostriker1992} Cen R., Ostriker J.~P., 1992, ApJ, 
399, L113 

\bibitem[\protect\citeauthoryear{de Blok \& 
Walter}{2006}]{Deblok&Walter2006} de Blok W.~J.~G., Walter F., 2006, AJ, 
131, 363 

\bibitem[\protect\citeauthoryear{Elmegreen}{2002}]{Elmegreen2002} 
Elmegreen B.~G., 2002, ApJ, 577, 206 

\bibitem[\protect\citeauthoryear{Elmegreen \& 
Parravano}{1994}]{Elmegreen&Parravano1994} Elmegreen B.~G., Parravano
  A., 1994, ApJ, 435, L121 

\bibitem[Ferland(2000)]{Ferland2000} 
Ferland, G.~J.\ 2000, Revista Mexicana de Astronomia y Astrofisica
Conference Series, 9, 153

\bibitem[\protect\citeauthoryear{Gerritsen \& 
Icke}{1997}]{Gerritsen&Icke1997} Gerritsen J.~P.~E., Icke V., 1997, A\&A, 
325, 972 

\bibitem[\protect\citeauthoryear{Gnedin}{1996}]{Gnedin1996} Gnedin 
N.~Y., 1996, ApJ, 456, 1 

\bibitem[\protect\citeauthoryear{Guiderdoni}{1987}]{Guiderdoni1987} 
Guiderdoni B., 1987, A\&A, 172, 27 

\bibitem[Haardt \& Madau(2001)]{Haardt&Madau2001}
Haardt, F., \& Madau, P. 2001, to be published in the proceedings of
XXXVI Rencontres de Moriond, astro-ph/0106018

\bibitem[\protect\citeauthoryear{Hernquist}{1990}]{Hernquist1990} 
Hernquist L., 1990, ApJ, 356, 359

\bibitem[\protect\citeauthoryear{Heyer et al.}{2004}]{Heyer2004} 
Heyer M.~H., Corbelli E., Schneider S.~E., Young J.~S., 2004, ApJ,
602, 723 

\bibitem[\protect\citeauthoryear{Katz}{1992}]{Katz1992} Katz N., 
1992, ApJ, 391, 502 

\bibitem[\protect\citeauthoryear{Katz, Weinberg, \& 
Hernquist}{1996}]{Katz1996} Katz N., Weinberg D.~H., Hernquist 
L., 1996, ApJS, 105, 19 

\bibitem[\protect\citeauthoryear{Kauffmann, White, \& 
Guiderdoni}{1993}]{Kauffmann1993} Kauffmann G., White S.~D.~M., 
Guiderdoni B., 1993, MNRAS, 264, 201 

\bibitem[\protect\citeauthoryear{Kawata \& 
Gibson}{2003}]{Kawata&Gibson2003} Kawata D., Gibson B.~K., 2003, MNRAS, 
340, 908 

\bibitem[\protect\citeauthoryear{Kay et al.}{2002}]{Kay2002} 
Kay S.~T., Pearce F.~R., Frenk C.~S., Jenkins A., 2002, MNRAS, 330,
113 
\bibitem[\protect\citeauthoryear{Kennicutt}{1989}]{Kennicutt1989} 
Kennicutt R.~C., Jr., 1989, ApJ, 344, 685 

\bibitem[\protect\citeauthoryear{Kennicutt}{1998a}]{Kennicutt1998review} 
Kennicutt R.~C., Jr., 1998a, ARA\&A, 36, 189 

\bibitem[\protect\citeauthoryear{Kennicutt}{1998b}]{Kennicutt1998} 
Kennicutt R.~C., Jr., 1998b, ApJ, 498, 541 

\bibitem[\protect\citeauthoryear{Kennicutt et 
al.}{2007}]{Kennicutt2007} Kennicutt R.~C., Jr., et al., 2007, arXiv, 
708, arXiv:0708.0922 

\bibitem[\protect\citeauthoryear{Komugi et al.}{2005}]{Komugi2005} 
Komugi S., Sofue Y., Nakanishi H., Onodera S., Egusa F., 2005, PASJ, 57, 
733 

\bibitem[\protect\citeauthoryear{Kravtsov}{2003}]{Kravtsov2003} 
Kravtsov A.~V., 2003, ApJ, 590, L1 

\bibitem[\protect\citeauthoryear{Krumholz \& 
McKee}{2005}]{Krumholz&McKee2005} Krumholz M.~R., McKee C.~F., 2005, ApJ, 
630, 250 

\bibitem[\protect\citeauthoryear{Li et al.}{2005}]{Li2005} Li Y., Mac
  Low M.-M., Klessen R.~S.,  
2005, ApJ, 626, 823 

\bibitem[\protect\citeauthoryear{Li, Mac Low, \& 
Klessen}{2006}]{Li2006} Li Y., Mac Low M.-M., Klessen R.~S., 
2006, ApJ, 639, 879 

\bibitem[\protect\citeauthoryear{Marri \& 
White}{2003}]{Marri&White2003} Marri S., White S.~D.~M., 2003, MNRAS, 
345, 561 

\bibitem[\protect\citeauthoryear{Martin \& 
Kennicutt}{2001}]{Martin&Kennicutt2001} Martin C.~L., Kennicutt R.~C., Jr., 
2001, ApJ, 555, 301 

\bibitem[\protect\citeauthoryear{Maybhate et 
al.}{2007}]{Maybate2007} Maybhate A., Masiero J., Hibbard J.~E., 
Charlton J.~C., Palma C., Knierman K.~A., English J., 2007, arXiv, 707, 
arXiv:0707.3582 

\bibitem[\protect\citeauthoryear{Meurer et al.}{1996}]{Meurer1996}
Meurer G.~R., Carignan C., Beaulieu S.~F., Freeman K.~C., 1996, AJ, 111, 
1551 

\bibitem[\protect\citeauthoryear{Mihos \& 
Hernquist}{1994}]{Mihos&Hernquist1994} Mihos J.~C., Hernquist L., 1994, 
ApJ, 437, 611 

\bibitem[\protect\citeauthoryear{Mo, Mao, \& 
White}{1998}]{MoMaoWhite1998} Mo H.~J., Mao S., White S.~D.~M., 1998, 
MNRAS, 295, 319 

\bibitem[\protect\citeauthoryear{Olling}{1995}]{Olling1995} Olling 
R.~P., 1995, AJ, 110, 591 

\bibitem[\protect\citeauthoryear{Navarro \& 
White}{1993}]{Navarro&White1993} Navarro J.~F., White S.~D.~M., 1993, 
MNRAS, 265, 271 

\bibitem[\protect\citeauthoryear{Navarro, Frenk, \& 
White}{1996}]{NFW} Navarro J.~F., Frenk C.~S., White 
S.~D.~M., 1996, ApJ, 462, 563 

\bibitem[\protect\citeauthoryear{Okamoto et 
al.}{2003}]{Okamoto2003} Okamoto T., Jenkins A., Eke V.~R., Quilis 
V., Frenk C.~S., 2003, MNRAS, 345, 429 

\bibitem[\protect\citeauthoryear{Quirk}{1972}]{Quirk1972} Quirk 
W.~J., 1972, ApJ, 176, L9 

\bibitem[\protect\citeauthoryear{Schaye}{2001a}]{Schaye2001} Schaye 
J., 2001a, ApJ, 559, 507 

\bibitem[\protect\citeauthoryear{Schaye}{2001b}]{Schaye2001maxNHI} Schaye 
J., 2001b, ApJ, 562, L95 

\bibitem[\protect\citeauthoryear{Schaye}{2004}]{Schaye2004} Schaye 
J., 2004, ApJ, 609, 667 (S04)

\bibitem[\protect\citeauthoryear{Schaye}{2007}]{Schaye2007} Schaye,
  J.\ 2007, in IAU Symp.\ 244, arXiv:0708.3366 

\bibitem[\protect\citeauthoryear{Schmidt}{1959}]{Schmidt1959} 
Schmidt M., 1959, ApJ, 129, 243 

\bibitem[\protect\citeauthoryear{Schuster et 
al.}{2007}]{Schuster2007} Schuster K.~F., Kramer C., Hitschfeld M., 
Garcia-Burillo S., Mookerjea B., 2007, A\&A, 461, 143 

\bibitem[\protect\citeauthoryear{Silk}{1997}]{Silk1997} Silk J., 
1997, ApJ, 481, 703 

\bibitem[\protect\citeauthoryear{Skillman}{1987}]{Skillman1987} 
Skillman E.~D., 1987, sfig.conf, 263 

\bibitem[\protect\citeauthoryear{Sommer-Larsen, G{\"o}tz, \& 
Portinari}{2003}]{Sommer-Larsen2003} Sommer-Larsen J., G{\"o}tz M., 
Portinari L., 2003, ApJ, 596, 47 

\bibitem[\protect\citeauthoryear{Springel}{2000}]{Springel2000} 
Springel V., 2000, MNRAS, 312, 859 

\bibitem[\protect\citeauthoryear{Springel, Yoshida, \& 
White}{2001}]{Springel2001} Springel V., Yoshida N., White S.~D.~M., 
2001, NewA, 6, 79 

\bibitem[\protect\citeauthoryear{Springel \& 
Hernquist}{2003}]{Springel&Hernquist2003} Springel V., Hernquist L., 2003, 
MNRAS, 339, 289 

\bibitem[\protect\citeauthoryear{Springel}{2005}]{Springel2005} 
Springel V., 2005, MNRAS, 364, 1105 

\bibitem[\protect\citeauthoryear{Springel, Di Matteo, \& 
Hernquist}{2005}]{Springeletal2005} Springel V., Di Matteo T., Hernquist 
L., 2005, MNRAS, 361, 776 

\bibitem[\protect\citeauthoryear{Steinmetz \& 
Mueller}{1994}]{Steinmetz&Mueller1994} Steinmetz M., Mueller E., 1994, A\&A, 
281, L97 

\bibitem[\protect\citeauthoryear{Summers}{1993}]{Summers1993} 
Summers F.~J., 1993, PhD thesis, Univ.\ California
 
\bibitem[\protect\citeauthoryear{Tan}{2000}]{Tan2000} Tan 
J.~C., 2000, ApJ, 536, 173 

\bibitem[\protect\citeauthoryear{Tasker \& 
Bryan}{2006}]{Tasker&Bryan2006} Tasker E.~J., Bryan G.~L., 2006,
  ApJ, 641, 878 

\bibitem[\protect\citeauthoryear{Thacker \& 
Couchman}{2000}]{Thacker&Couchman2000} Thacker R.~J., Couchman H.~M.~P., 
2000, ApJ, 545, 728 

\bibitem[\protect\citeauthoryear{Toomre}{1964}]{Toomre1964} Toomre 
A., 1964, ApJ, 139, 1217 

\bibitem[\protect\citeauthoryear{Tutukov}{2006}]{Tutukov2006} 
Tutukov A.~V., 2006, ARep, 50, 526 

\bibitem[\protect\citeauthoryear{Wong \& Blitz}{2002}]{Wong&Blitz2002} 
Wong T., Blitz L., 2002, ApJ, 569, 157 

\bibitem[\protect\citeauthoryear{Yepes et al.}{1997}]{Yepes1997} 
Yepes G., Kates R., Khokhlov A., Klypin A., 1997, MNRAS, 284, 235 

\bibitem[\protect\citeauthoryear{Zhang, Fall, \& 
Whitmore}{2001}]{Zhang2001} Zhang Q., Fall S.~M., Whitmore B.~C., 
2001, ApJ, 561, 727 

\end{thebibliography}
\end{document}